\documentclass[useAMS, fleqn,usenatbib]{mnras}

\usepackage{newtxtext,newtxmath}
\usepackage{multirow}
\usepackage{textcomp, gensymb}  
\usepackage{siunitx}
\usepackage{xspace}
\usepackage{lineno}
\usepackage{ulem}
\usepackage{soul}
\usepackage{changes}

\usepackage[T1]{fontenc}

\newcommand{\kms}{${\rm km\cdot s^{-1}}$}
\newcommand{\thestar}{Feige 64}
\newcommand{\comp}{_\textrm{comp}}

\DeclareRobustCommand{\VAN}[3]{#2}
\let\VANthebibliography\thebibliography
\def\thebibliography{\DeclareRobustCommand{\VAN}[3]{##3}\VANthebibliography}

\usepackage{graphicx}	
\usepackage{amsmath}	

\title[A Helium-shell Burning BHB Star from CEE]{A Helium-shell Burning Blue Horizontal Branch Star Produced from Common Envelope Evolution}

\author[Jiao Li et al.]{
Jiao Li,$^{1,2,3}$\thanks{Contact e-mail: lijiao@ynao.ac.cn;  liuchao@nao.cas.cn}
Changqing Luo,$^{2}$
Hai-Liang Chen,$^{1, 3}$
Zhicun Liu,$^{4}$
Bo Zhang,$^{2}$
Shi Jia,$^{1,3}$
Hongwei Ge,$^{1,3}$
\and Tao Wu,$^{1,3}$
Yuhan Yao,$^{5,6}$
Pei Wang,$^{2}$
Marat Gilfanov,$^{7,8}$
You Wu,$^{2}$
Zhenwei Li,$^{1,3}$
Zhengwei Liu,$^{1,3}$
\and Xiangcun Meng,$^{1,3}$
Xue-Fei Chen,$^{1,3}$
Philipp Podsiadlowski,$^{9,10,11}$
Chao Liu,$^{2,12}$
and Zhan-Wen Han,$^{1,3}$
\\
$^{1}$International Centre of Supernovae (ICESUN), Yunnan Key Laboratory of Supernova Research, Yunnan Observatories, Chinese Academy of Sciences,\\ Kunming 650216, People's Republic of China\\
$^{2}$National Astronomical Observatories, Chinese Academy of Sciences, Beijing, 100101, China\\
$^{3}$Key Laboratory for the Structure and Evolution of Celestial Objects, CAS, Kunming 650216, China\\
$^4$Department of Physics, Hebei Normal University, Shijiazhuang 050024, People's Republic of China\\
$^5$Miller Institute for Basic Research in Science, 206B Stanley Hall, Berkeley, CA 94720, USA\\
$^6$Department of Astronomy, University of California, Berkeley, CA 94720-3411, USA\\
$^7$Space Research Institute, Russian Academy of Sciences, Profsoyuznaya 84/32, 117997 Moscow, Russia\\
$^8$Max Planck Institute for Astrophysics, Karl-Schwarzschild-Str 1, Garching b. München D-85741, Germany\\
$^9$London Centre for Stellar Astrophysics, Vauxhall, London, United Kingdom\\
$^{10}$University of Oxford, St Edmund Hall, Oxford, OX1 4AR, United Kingdom\\
$^{11}$Heidelberger Institut f{\"u}r Theoretische Studien, Schloss-Wolfsbrunnenweg 35, 69118 Heidelberg, Germany\\
$^{12}$University of Chinese Academy of Sciences, Beijing, 100049, China
}

\date{Accepted XXX. Received YYY; in original form ZZZ}
\date{Accepted 2026 August 11. Received 2026 August 10; in original form 2025 December 09}

\pubyear{\the\year{}}

\begin{document}
\label{firstpage}
\pagerange{\pageref{firstpage}--\pageref{lastpage}}
\maketitle

\begin{abstract}
Observationally, blue horizontal branch (BHB) stars are defined as hot stars occupying a characteristic region between the extreme blue horizontal branch and RR Lyrae variables in the Hertzsprung–Russell diagram. Most of them are interpreted as stripped core-helium-burning stars, but the role of binary interaction in their formation remains unclear. Here, we report the discovery of a metal-rich BHB star in a 0.82628-day binary system (\thestar) comprising a $0.35\pm0.03\,M_{\odot}$ BHB star and a likely $1.26\pm0.17\,M_{\odot}$ white dwarf (WD). The BHB star has an effective temperature of $15{,}524\pm310$\, K and a luminosity of $39.7\pm4.1\,L_{\odot}$. Stellar evolution modelling indicates that it is a helium-shell-burning star produced through the common-envelope channel, retaining a hydrogen-rich envelope that is more massive than previously thought for low-mass stars. This finding provides direct evidence for binary interaction in the formation of BHB stars, offering a fresh perspective on interpreting this emerging population.
\end{abstract}




\section{Introduction}\label{sec:intro}

BHB stars are evolved, core-helium-burning stars with thin hydrogen-rich envelopes, typically occupying the region between RR Lyrae variables and extreme horizontal branch stars in the Hertzsprung–Russell diagram \citep[e.g.,][]{Catelan2009Ap&SS.320..261C}. Since the early photometric studies of globular clusters, these stars have served as vital probes for understanding stellar evolution and Galactic structure.

The foundational mapping of the horizontal branch (HB) began with the work of \cite{Sandage1953AJ.....58...61S}, who provided some of the first detailed color-magnitude diagrams (CMDs) of globular clusters like M3, identifying the horizontal sequence of stars extending from the red giant branch toward higher temperatures. As spectroscopic capabilities advanced, investigations of faint blue stars in the Galactic field were carried out \citep[e.g.,][]{Greenstein1966ApJ...144..496G, Greenstein1974ApJS...28..157G}, revealing a population of low-luminosity objects inconsistent with the standard main sequence. These observations bridged the gap between HB stars in globular clusters and their counterparts in the general Galactic field.

The formation mechanisms of BHB stars, particularly those in the Galactic field, remain a subject of debate, centering on the mass-loss processes required to strip their hydrogen envelopes. To reach the blue end of HB, a star must lose a substantial fraction of its hydrogen-rich envelope during the red giant branch (RGB) phase. \cite{Newell1973ApJS...26...37N} highlighted the morphological complexity of the HB, identifying distinct “gaps” in the distribution of HB stars that imply discrete physical mechanisms or evolutionary “bins” governing their effective temperatures and luminosities. Subsequent theoretical work by \cite{Sweigart1997ApJ...474L..23S} further demonstrated how variations in helium abundance, stellar rotation, and mass-loss efficiency can drive stars toward these high-temperature configurations. In addition, several types of binary interactions have also been proposed as potential formation channels\citep[e.g.,][]{Leizhenxin2013A&A...549A.145L, Bobrick2024MNRAS.52712196B}.

With the advent of large spectroscopic surveys and precise Gaia parallaxes, metal-rich field BHB candidates have been uncovered during BHB identification efforts \citep[e.g.,][]{Jujie2024ApJS..270...11J}. The properties of these objects are difficult to reconcile with single-star evolution alone. Binary interactions are therefore expected to play a crucial role in producing metal-rich BHB stars, analogous to the binary formation channels that yield hot subdwarfs (sdOB; \citealt{Hanzhanwen2002MNRAS.336..449H, Geier2022A&A...661A.113G}). In particular, the common-envelope ejection (CEE) process can strip the envelope of a red giant, exposing its helium core and leaving behind a close binary system. However, direct evidence that such a channel can produce BHB-like stars—retaining a substantial hydrogen envelope—has been lacking.


In this paper, we present \thestar{}, a system formed through CEE, and show that it appears as a BHB star undergoing helium-shell burning. Section~\ref{sec:obsevation} describes the observations and data reduction, including optical spectroscopy, photometry, and radio pulsation searches. Section~\ref{sec:analysis} presents the determination of the physical parameters. Section~\ref{sec:theory} discusses the nature and formation history of \thestar{}. Section~\ref{sec:conclusion} summarizes our main results.

\section{Observation and Data}\label{sec:obsevation}

\begin{figure}
    \centering
    \includegraphics[width=\columnwidth]{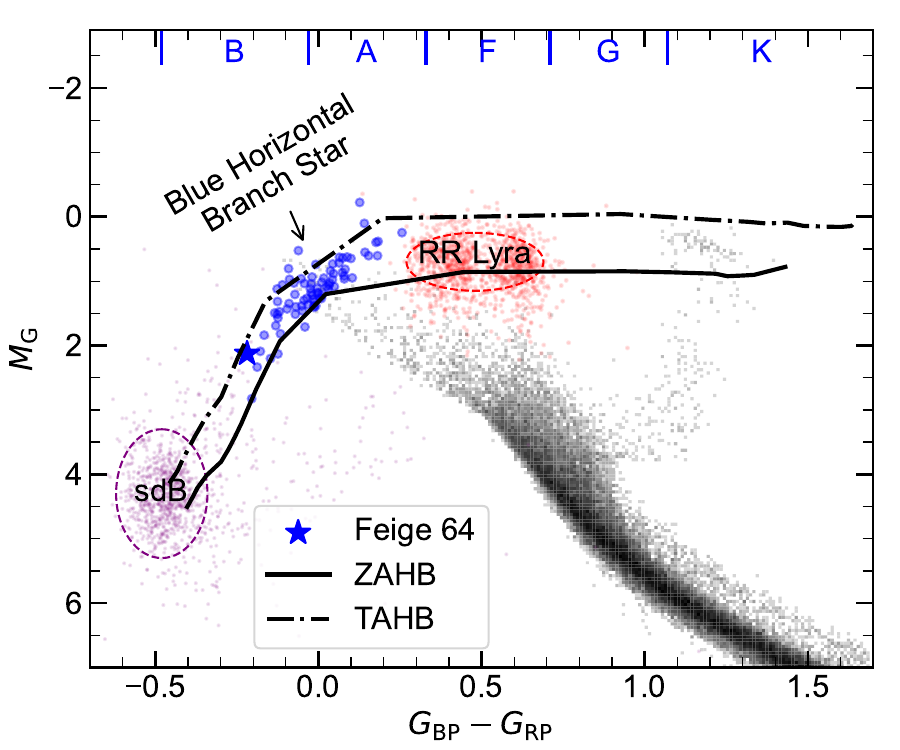}
    \caption{{\it Gaia} $G$, $G_{\rm BP}-G_{\rm RP}$ color-magnitude diagram for Feige 64.
    The HB band (limited by zero-age HB, ZAHB and the terminal-age HB, TAHB) with [Fe/H] = 0 is taken from \citet{Dorman1993ApJ...419..596D}.
    The blue dots are BHBs of the NGC 6397 and 6752 globular clusters (the selection criteria of the members of two clusters are in Appendix~\ref{sec:GC_members}).
    The purple dots are sdBs compiled by \citet{Geier2020A&A...635A.193G} with parallax\_over\_error $>$ 5.
    The red dots are a fraction of RR Lyra variables downloaded from {\it Gaia} DR3. 
    The color and magnitude of the stars were corrected using the dust reddening map of \citet{Schlegel1998ApJ...500..525S} or the 3D dust map implanted in {\tt dustmaps}~\citep{Green2019ApJ...887...93G}. The color excess $E(G_{\rm BP}-G_{\rm RP})$ and the extinction $A_G$ are calculated with $E(B-V)$ and the relative extinction values from Table 3 of \citet{Wangshu2019ApJ}. The gray background is the density diagram of the solar neighbor stars within 100 pc selected from {\it Gaia} DR2.}
    \label{fig:cmd}
\end{figure}

Since 2020, we have conducted a radial-velocity monitoring campaign of BHB and sdB candidates selected from the Large Sky Area Multi-Object Fiber Spectroscopic Telescope (LAMOST; \citealt{Cui2012RAA....12.1197C}) survey, aiming to probe the formation pathways of metal-rich BHB stars. One of the targets, \thestar{}, is a previously known faint blue star \citep{Feige1958ApJ...128..267F, Kilkenny1977MNRAS.181..611K}. Its photometric variability has been detected by the Asteroid Terrestrial-impact Last Alert System (ATLAS; \citealt{Heinze2018AJ....156..241H}) and \textit{Gaia}, with \textit{Gaia} DR3 classifying it as an eclipsing-binary candidate \citep{Distefano2023A&A...674A..20D, Mowlavi2023A&A...674A..16M}; sinusoidal variability was also reported from \textit{Gaia} DR3 multi-epoch photometry \citep{Ranaivomanana2025A&A...693A.268R}. By combining the  Spectroscopic measurements with light curves from the \textit{Transiting Exoplanet Survey Satellite} (\textit{TESS}; \citealt{Ricker2015JATIS...1a4003R}), we characterize \thestar{} as an ellipsoidal variable in a close binary system containing a metal-rich BHB component (Fig.~\ref{fig:cmd}).

\subsection{Spectroscopic and radial velocity}\label{sec:spec}

We conducted multi-epoch observations of \thestar{} using the Palomar 200-inch (P200/DBSP), Keck I (HIRES), and Xinglong 216-cm (X216/BFOSC) telescopes. P200/DBSP spectra ($R\sim2500$) were obtained in January 2020 covering 3750–5250 \AA, while high-resolution ($R\sim30000$) spectra were acquired with Keck-I/HIRES in February 2020. Additionally, 18 spectra ($R\sim2000$) were collected using X216/BFOSC between March 2020 and May 2021. Data reduction—including bias subtraction, flat-fielding, and wavelength calibration—was performed using standard IRAF packages\citep{iraf1986SPIE..627..733T, iraf1993ASPC...52..173T} and our in-house Python tool \href{https://github.com/lidihei/pyexspec}{\tt pyexspec}.

Radial velocities (RVs) were determined by cross-correlating the observed spectra with a grid of TLUSTY B-type synthetic templates \citep{Lanz2007ApJS..169...83L}, spanning $15000 \leq T_{\rm eff} \leq 17000$ K and $3.0 \leq \log g \leq 4.25$. Synthetic templates were degraded to match the respective instrumental resolutions ($R \approx 30000, 2500, 2000$) using Gaussian kernels. RV measurements and uncertainty estimation were performed using 100 Monte Carlo (MC) samples with the {\tt Laspec} package \citep{Zhangbo2020ApJS..246....9Z}. All resulting radial velocities are provided in Table~\ref{tab:rv}.

\begin{table*}
\centering
\begin{tabular}{ccccccccc}
\hline
Facility     & UT shut   & BJD\footnotesize{mid} & Exposure & RV & Phase &SNR\\
n.a.   &     yyyy-mm-dd hh:mm:ss &day&s&${\rm kms^{-1}}$  &  n.a.&n.a.\\
\hline
\multirow{3}{*}{P200/DBSP}
& 2020-01-13T10:35:43 & 2458861.951806 & 1200 & $131.17\pm0.61$ & -0.114 & 150.2\\
& 2020-01-13T10:56:05 & 2458861.965947 & 1200 & $116.79\pm0.63$ & -0.097 & 154.1\\
& 2020-01-13T11:16:27 & 2458861.980087 & 1200 & $96.73\pm0.57$ & -0.080 & 156.3\\
\hline
\multirow{3}{*}{KECK/HIRES}
& 2020-02-05T14:05:56 & 2458885.093073 & 204 & $114.85\pm1.13$ & -0.108 & 34.4\\
& 2020-02-05T14:45:48 & 2458885.120728 & 200 & $79.35\pm0.99$ & -0.074 & 35.1\\
& 2020-02-05T16:03:31 & 2458885.174698 & 200 & $6.57\pm0.95$ & -0.009 & 33.9\\
\hline
\multirow{17}{*}{X216/E9+G10}
& 2020-03-14T16:34:29 & 2458923.205866 & 1800 & $-30.46\pm2.61$ & 0.018 & 42.3\\
& 2020-03-14T17:04:35 & 2458923.226768 & 1800 & $-59.24\pm2.67$ & 0.043 & 38.0\\
& 2020-03-15T17:29:07 & 2458924.243792 & 1800 & $-215.32\pm14.02$ & 0.274 & 21.6\\
& 2020-03-15T17:59:13 & 2458924.264694 & 1800 & $-212.69\pm12.38$ & 0.300 & 22.1\\
& 2020-03-17T16:07:22 & 2458926.186992 & 1800 & $114.99\pm5.38$ & -0.374 & 26.5\\
& 2020-03-21T20:23:42 & 2458930.364923 & 1800 & $158.27\pm5.91$ & -0.318 & 26.3\\
& 2020-03-22T18:15:35 & 2458931.275934 & 1800 & $169.68\pm2.40$ & -0.215 & 42.5\\
& 2020-03-27T20:19:19 & 2458936.361732 & 1800 & $52.01\pm5.61$ & -0.060 & 27.8\\
& 2020-03-28T20:30:21 & 2458937.369365 & 1800 & $-173.51\pm4.42$ & 0.159 & 31.6\\
& 2020-04-02T19:49:10 & 2458942.340608 & 1800 & $-160.89\pm3.96$ & 0.176 & 36.9\\
& 2021-03-21T19:21:37 & 2459295.321825 & 1800 & $-126.51\pm5.38$ & 0.370 & 26.6\\
& 2021-05-04T15:45:05 & 2459339.169573 & 1800 & $-64.85\pm4.37$ & 0.437 & 26.1\\
& 2021-05-04T17:36:51 & 2459339.247184 & 1800 & $29.07\pm6.17$ & -0.469 & 25.6\\
& 2021-05-05T12:00:14 & 2459340.013375 & 1800 & $-52.12\pm10.77$ & 0.458 & 22.9\\
& 2021-05-05T13:12:30 & 2459340.063557 & 1800 & $19.88\pm6.48$ & -0.481 & 28.5\\
& 2021-05-08T13:33:47 & 2459343.078150 & 1800 & $-180.12\pm3.48$ & 0.167 & 33.5\\
& 2021-05-08T16:11:17 & 2459343.187518 & 1800 & $-175.46\pm4.04$ & 0.299 & 33.7\\
\hline
\end{tabular}
\caption{Radial velocity information.
}
\label{tab:rv}
\end{table*}

\subsection{Photometric data and period analysis}\label{sec:tesslc}

\thestar{} was observed by \textit{TESS} during Sector~22
(February~19 to March~17, 2020) with a 30-min cadence.
The data were obtained from the Mikulski Archive for Space Telescopes (MAST)
and reduced using the MIT Quick Look Pipeline (QLP;
\citealt{Huang2020RNAAS...4..204H}).
We adopted the normalized, Kepler-spline-detrended light curve
(KSPSAP\_FLUX) and excluded data points with normalized fluxes below 0.9.

\begin{figure}
    \centering
    \includegraphics[width=1\columnwidth]{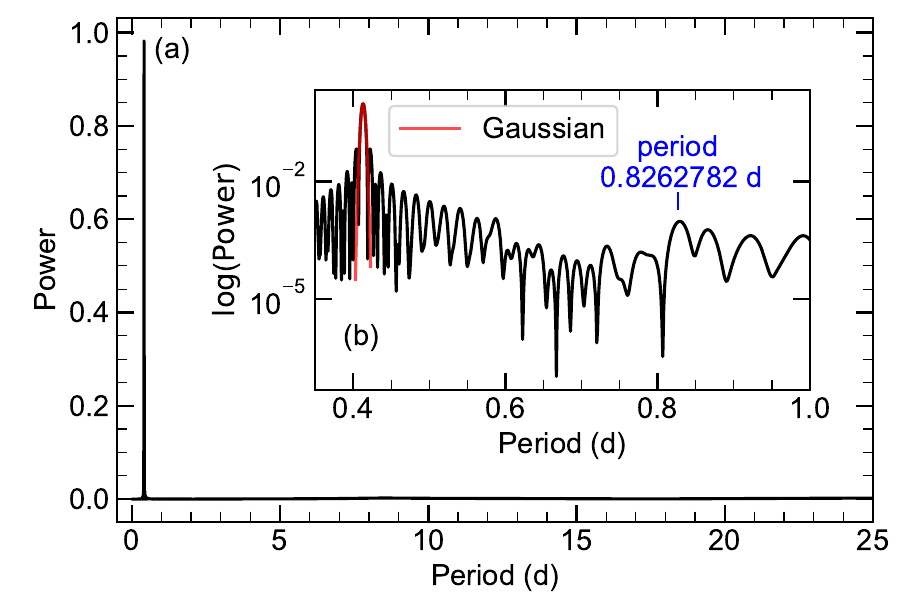}
    \caption{The Lomb–Scargle periodogram of the TESS light curve. In the full periodogram, a dominant peak can be seen at 0.4131391(55)
    days corresponding to half of the orbital period of 0.8262782(110). \textbf{Panel (b)}: the red line shows the best-fitted Gaussian function using non-linear least squares to fit the periodogram data around the peak period. The periodogram power is shown on a logarithmic scale for clarity. The error is the square root of the estimated covariance. The peak at the orbital period is too weak to be unambiguously identified, as the height difference between the two maxima of the light curve is merely $\sim 1$\% of the overall amplitude, see Fig.~\ref{fig:lcrvfit} (a).}
    \label{fig:Lomb-Scargle}
\end{figure}

\begin{figure*}
    \centering
    \includegraphics[scale=0.67, angle=0]{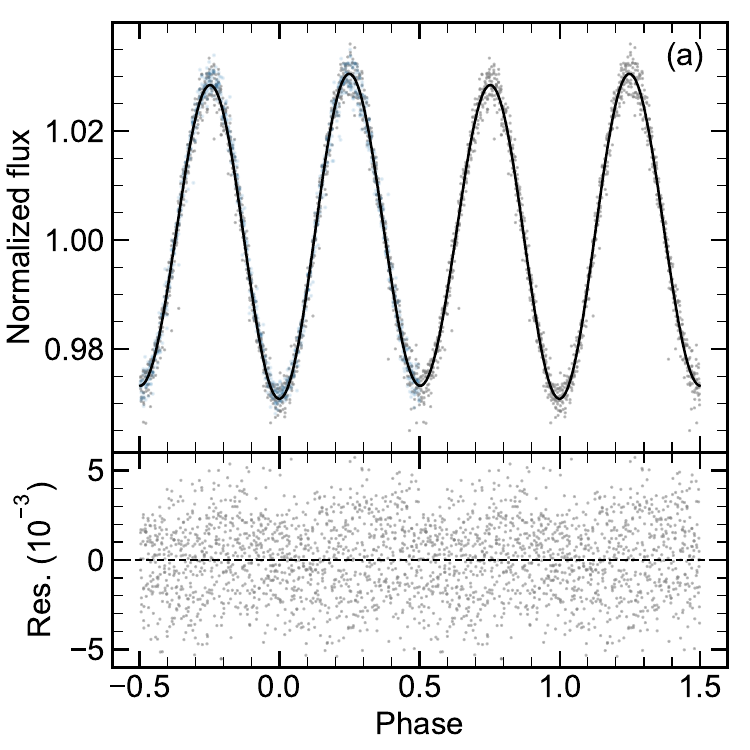}
    \includegraphics[scale=0.67, angle=0]{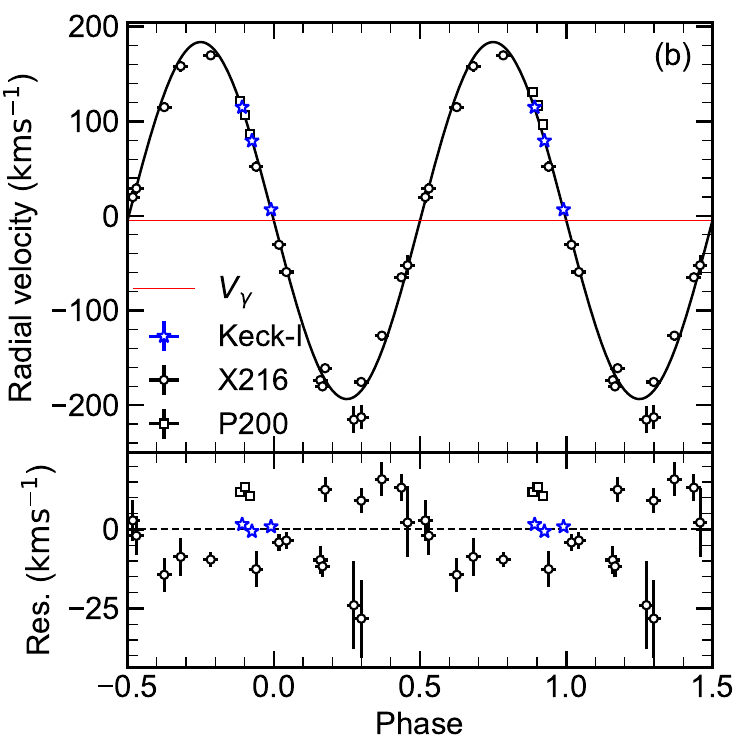}
    \caption{Phased data for \thestar{}. \textbf{(a)}: Phase-folded TESS light curve and the fitting residual. The small gray dots are the original observation flux of TESS. The black line represents the best-fit model generated by {\tt LCURVE}. \textbf{(b)}:  Phase-folded radial velocity curve and fitting residual. The black stars, dots, and squares are radial velocities measured from spectra observed by Keck-I,  Xinglong 216-cm, and Palomar 200-inch telescopes, respectively. The red line is the systematic radial velocity of the binary.
    {\it Notes :} The folded curves are just duplicated into two orbital cycles for clarity.
    }
    \label{fig:lcrvfit}
\end{figure*}

A Lomb--Scargle periodogram of the \textit{TESS} light curve reveals a
highly significant peak at a period of $0.4131391(55)$~d
(Fig.~\ref{fig:Lomb-Scargle}).
The period search was performed using the \texttt{LombScargle} implementation
in \texttt{Astropy} \citep{astropy:2013}.
When folding the light curve and the radial-velocity measurements at twice
this period, two maxima are observed in the light curve and a single maximum
in the radial-velocity curve (Fig.~\ref{fig:lcrvfit}),
consistent with an ellipsoidal modulation.
We therefore adopt an orbital period of $0.8262782(110)$~d, in agreement with
the period reported in the first \textit{Gaia} catalogue of eclipsing-binary
candidates \citep{Mowlavi2023A&A...674A..16M}.

\begin{figure}
    \centering
    \includegraphics[width=1\columnwidth]{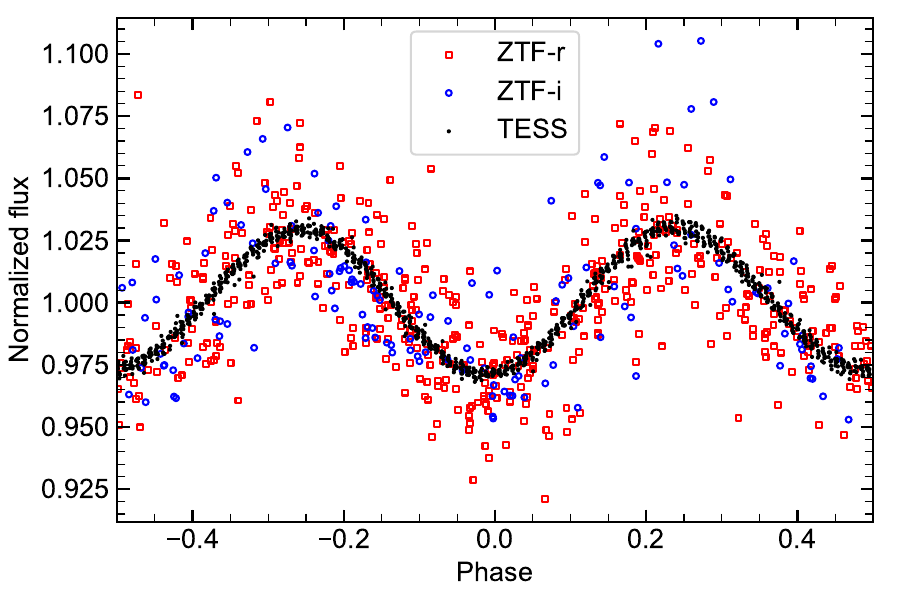}
    \caption{Phase folded TESS and ZTF light curves. The flux is normalized by dividing the median value. The square, circle, and dot are the ZTF $r$-band, $i$-band, and the corrected TESS light curves, respectively. The folding period is 0.8262782 days.}
    \label{fig:tess_ztf}
\end{figure}

A faint nearby source located $8.4\arcsec$ away may dilute the \textit{TESS}
photometry given its large pixel size ($21\arcsec$).
The neighbor shows no variability at the orbital period in the
Zwicky Transient Facility (ZTF) data \citep{Masci2019PASP..131a8003}.
So, we corrected the \textit{TESS} light curve for third-light contamination by following the formula,
\begin{equation}
    \tilde{y}(t) = \frac{y(t)-1+f}{f},
\end{equation}\label{eq:correctlc}
where $y(t)$ is the normalized flux by dividing the median flux of the light curve. $\tilde{y}(t)$ is the corrected normalized flux. $f$ is the flux ratio of the object's median flux to the median flux going through the photometry aperture,
\begin{equation}
    f = \frac{F_{\rm obj}}{F_{\rm obj}+F_{\rm nb}} = \frac{1}{1+10^{(m_{\rm obj} - m_{\rm nb})/2.5}},
\end{equation}\label{eq:fluxratio}
where $F_{\rm obj}$ and $F_{\rm nb}$ are median fluxes of object and nearby star, respectively; $m_{\rm obj}$ and $m_{\rm nb}$ are apparent magnitudes of object and nearby star, respectively.
We approximately obtained $f=0.974$ using the magnitudes of the {\it Gaia} $G_{\rm RP}-$band of the two sources ($m_{\rm obj} = 12.5737$ mag, $m_{\rm nb} = 16.5253$ mag), because their isolated TESS fluxes are unknown and the transmission curve of the TESS filter is very similar to that of the $G_{\rm RP}$-band. The corrected TESS light curve is in excellent agreement with both ZTF $r$-, and $i$-band light curves (Fig.~\ref{fig:tess_ztf}).

\subsection{Radio observations with FAST}\label{sec:fast}

We conducted targeted radio follow-up observations of \thestar{}
with the Five-hundred-meter Aperture Spherical radio Telescope (FAST) to search for potential pulsar emission from an unseen compact companion. Four observing sessions were carried out between January and April 2021, with a total on-source time of 1.75~h.
The observations were performed with the central beam of the 19-beam L-band receiver, covering 1.05--1.45~GHz with a system temperature of $\sim$25~K \citep{Jiang2020RAA....20...64J}. Data were recorded in pulsar search mode and stored in PSRFITS format \citep{Hotan2004PASA...21..302H}.

The data were searched for both periodic pulsations and single radio bursts using standard pulsar-search pipelines based on \texttt{PRESTO} \citep{Ransom2001PhDT.......123R, Wang2021SCPMA..6429562W} and \texttt{HEIMDALL} \citep{Champion2016MNRAS.460L..30C}. We performed de-dispersion over a dispersion-measure range of 0--200~pc~cm$^{-3}$ to account for uncertainties in Galactic electron-density models. No credible signals were detected in any of the observing sessions.

\section{Analysis}\label{sec:analysis}

\subsection{Spectroscopic atmospheric parameters}

We initially used spectra from P200/DBSP to measure the effective temperature, surface gravity, and metallicity. To ensure consistency, the observed spectra were adjusted to the rest frame. Quantitative spectral analysis was conducted using the non-local thermodynamic equilibrium (NLTE) line-blanketed model atmosphere grid for early B-type stars (BSTAR2006) from \sout{the} TLUSTY \citep{Lanz2007ApJS..169...83L}. The synthetic optical spectra were generated using {\tt Synspec} and can be downloaded from the TLUSTY website\footnote{\url{http://tlusty.oca.eu/Tlusty2002/tlusty-frames-BS06.html}}. 
We perform an MCMC approach to fit the normalized spectrum of DBSP with a posterior distribution,
\begin{equation}\label{eq:propspec}
    \ln[\mathcal{L}(\theta_{\rm spec} \mid f)] \propto \ln [\mathcal{L}(f \mid \theta_{\rm spec})]
    =-\frac{1}{2}\sum_{i=1}^{N}\frac{[f_{i}-f_{i}(\theta_{\rm spec})]^2}{\sigma^2_{f_i}}
\end{equation}
where $\theta_{\rm spec} = (T_{\rm eff}, \log g, Z, R_{\rm spec}, v\sin i, \epsilon)$; $T_{\rm eff}$, $\log g$, $Z$, $R_{\rm spec}$, $v\sin i$ and $\epsilon$ are effective temperature, surface gravity, metallicity abundance, spectral resolution of DBSP, the projected rotational velocity and limb-darkening coefficient, respectively. $i$ stands for the pixel of DBSP spectrum, and the pixels in wavelength of [3733, 3900] or [3950, 5200]~\AA\, are used. $f(\theta_{\rm spec})$ is the model spectrum. The model spectrum was obtained through the following process: 1) interpolation by using a Python package of \href{https://github.com/sczesla/PyAstronomy}{\tt regli} with the spectra grid of BSTAR2006,  2) broadening using both Gaussian kernel (for instrument broaden) and rotation kernel (described in Eq. 18.14 of {\it The Observation and Analysis of Stellar Photospheres} \citealt{gray_2021}), 3) resampling in the DBSP wavelength, 4) normalization.  The observed and model spectra were normalized using the same algorithm available in the Python package \href{https://github.com/hypergravity/laspec}{\tt Laspec} \citep{Zhangbo2020ApJS..246....9Z}. Here, we drop $v\sin i$ as it varies with $R_{\rm spec}$. We also noted that $\epsilon$ can not be constrained by DBSP spectrum. For each parameter, we calculated the average value from the measurements obtained from the three DBSP spectra and adopted this as the expected value. It is noted that the difference between the maximum and minimum values exceeds the error obtained by the MCMC approach (see Table~\ref{tab:spec_atom}). Therefore, we adopt this difference as the error estimate and obtain $T_{\rm eff} = 15524\pm 85$~ K, $\log g = 4.104\pm0.012$ (cgs), and $Z=1.67\pm0.15$~$Z_\odot$. 

\begin{figure*}
    \centering
    \includegraphics[scale=0.5, angle=0]{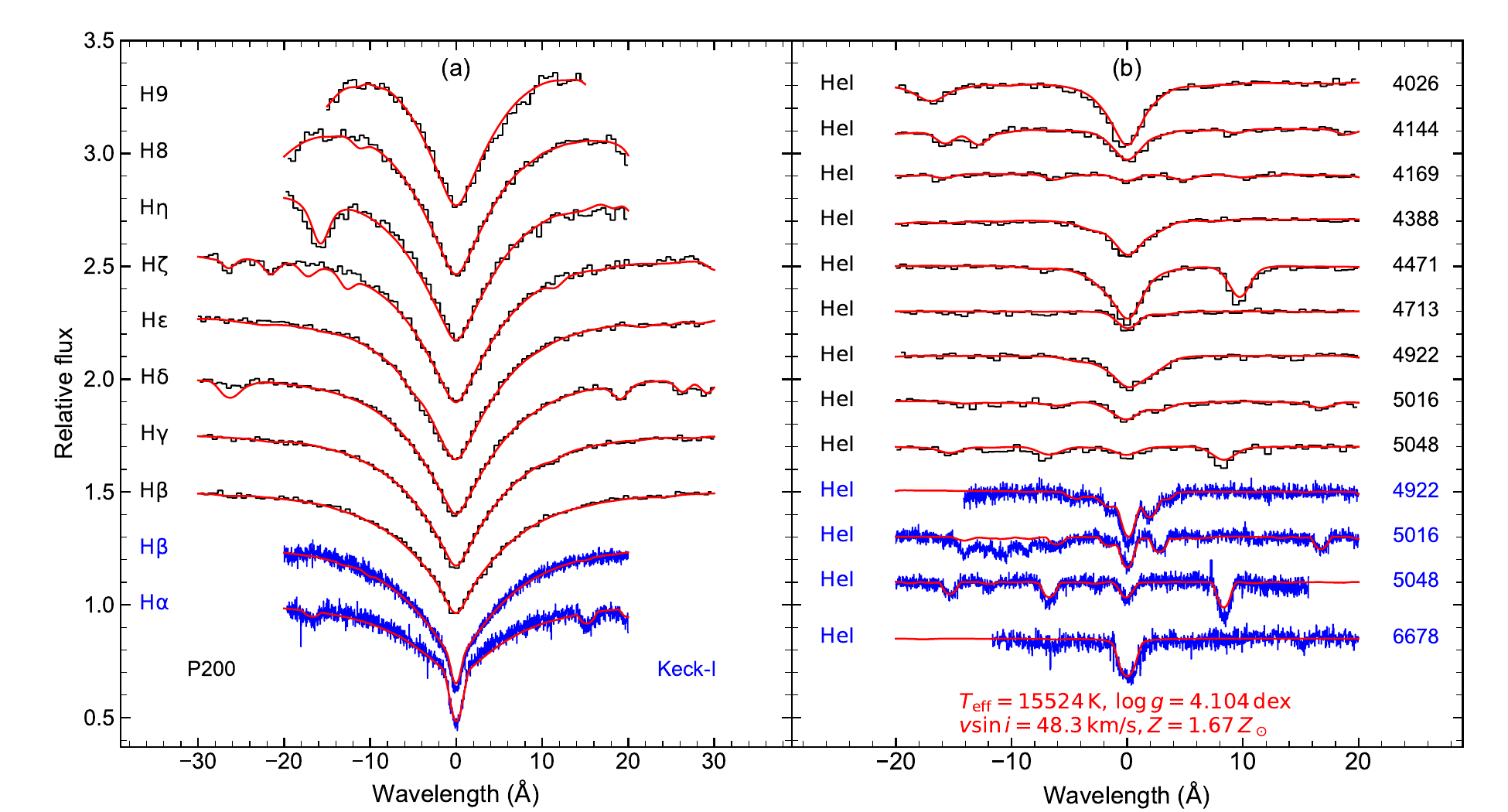}
    \caption{Panels \textbf{(a)} and \textbf{(b)} show a comparison between the hydrogen and helium lines in the model spectrum and those observed by P200 (2020-01-13T10:35:43) and Keck-I (2020-02-05T14:45:48) telescopes. Their wavelengths have been adjusted to the rest frame. The black and blue spectra correspond to observations taken by the P200 and Keck-I telescopes, respectively. The red lines represent the model spectrum interpolated from BSTAR2006 synthetic spectra, with atmosphere parameters $T_{\rm eff} = 15524$ K, $\log g = 4.104$ (cgs), and $Z=1.67$ $Z_\odot$. These lines are broadened with a rotational velocity of 48.3 \kms{} and degraded to match resolutions of P200 DBSP ($R=2500$) and Keck-I HIRES ($R=30000$).}
    \label{fig:specfit}
\end{figure*}

The microturbulence ($\xi$) of 8 \kms of \thestar{} is obtained by using the relationship between surface gravity and microturbulence \citep{Liuzhicun2022ApJ...937..110L}. 
To exclude the effect of macroturbulence, we also derived a projected rotational velocity $v\sin i = 48.3\pm 2$ \kms{} from He I 6678 \AA{} using the Fourier transform technique \citep{2007A&A...468.1063S, 2014A&A...562A.135S}, based on the radial-velocity-corrected co-added spectrum from three Keck-I exposures. We measured the helium, carbon, silicon, sulphur, and iron abundances of Feige 64 by using the interpolated solar-metallicity TLUSTY model atmospheres and SYNSPEC code, based on the above stellar parameters ($T_{\rm eff} = 15524\pm85$ K, $\log g = 4.104\pm 0.012$ (cgs) and $v\sin i = 48.3\pm 2$ \kms{}) and the co-added Keck-I spectrum (see Table~\ref{tab:abundanceVt=8}). The elemental abundances are determined by $\chi^2$ minimization fitting to selected individual lines. The abundance uncertainties include the contribution from the microturbulence uncertainty of 3 \kms{}.  Specifically, the total abundance error for each element is computed as $\sigma_{\rm All}$ = ($\sigma_{\rm A}^2$ + $\sigma_{\rm B}^2$)$^{1/2}$, where $\sigma_{\rm A}^2$ is the standard deviation of the abundances derived from all lines for the same element, and $\sigma_{\rm B}^2$ is the abundance variation induced by the microturbulence uncertainty (listed in Table~\ref{tab:specabundace}, while Fig.~\ref{fig:Compare} presents the best fits to the individual spectral lines of \thestar.) As Fig.~\ref{fig:specfit} shows, the hydrogen and helium lines of the model spectrum ($T_{\rm eff} = 15524$ K, $\log g = 4.104$ (cgs), and $v\sin i = 48.3$ \kms{}) are in good agreement with the observed spectra. Assuming a metal mass fraction of 0.018 for \thestar{}, the surface hydrogen and helium mass fractions were $0.63\pm0.12$ and $0.35\pm0.12$, respectively.

The metallicity ($Z=1.76\pm0.15~Z_\odot$) derived from the P200/DBSP spectra is presumably dominated by light elements in the optical spectrum (C, Si, S) and is less influenced by iron, which is sub-solar ($\rm [Fe/H]\sim -0.17$; see Table~\ref{tab:abundanceVt=8}). However, iron dominates the ultraviolet line blanketing and thus affects the atmospheric structure. We therefore adopt a fixed $Z=0.7~Z_\odot$ when fitting the P200/DBSP spectra, obtaining $T_{\rm eff}=15819\pm102$~K and $\log g = 4.104\pm0.029$ (cgs).

Compared with the previous solution, $\log g$ remains essentially unchanged but with a larger uncertainty, while $T_{\rm eff}$ differs by 295~K. To account for this systematic offset, we combine it in quadrature with the statistical uncertainty (85~K), resulting in a final effective temperature of $T_{\rm eff} = 15524 \pm 307$~K. We adopt $\log g = 4.10 \pm 0.03$ (cgs). These values are comparable to the \textit{Gaia}-based atmospheric parameters for hot stars, with $T_{\rm eff,\,esphs} = 15325\pm188$~K and $\log g_{\rm esphs} = 4.02\pm0.03$ (cgs) for \thestar, derived from combined BP/RP and RVS spectra \citep{Creevey2023A&A...674A..26C}.

\begin{table}
    \centering
    \begin{tabular}{ccccc}
    \hline
    Species & $\log[N(X)/N({\rm H})]$ & $\log[N(X)/N({\rm H})]_\odot$ & [$X/{\rm H}$] \\
    \hline
	He$^{\rm a}$ & $-0.89\pm0.25$ & $-1.07\pm0.01$ & $0.18\pm0.25$ \\
	C$^{\rm a}$ & $-3.33\pm0.09$ & $-3.57\pm0.05$ & $0.24\pm0.10$ \\
	Si$^{\rm a}$ & $-4.20\pm0.22$ & $-4.49\pm0.03$ & $0.29\pm0.22$ \\
	S$^{\rm a}$ & $-4.73\pm0.11$ & $-4.88\pm0.03$ & $0.15\pm0.11$ \\
	Fe$^{\rm b}$ & $-4.67\pm0.20$ & $-4.50\pm0.04$ & $-0.17\pm0.20$ \\
    \hline
    \multicolumn{4}{l}{$^{\rm a}$ represents the lines treated in NLTE.}\\
    \multicolumn{4}{l}{$^{\rm b}$ represents the lines treated in LTE.}\\
    \end{tabular}
    \caption{Elemental Abundances. The Solar abundances are from \citet{Asplund2009ARA&A..47..481A}. [X/H] =$\log[N(X)/N({\rm H})] - \log[N(X)/N({\rm H})]_\odot$
}
    \label{tab:abundanceVt=8}
\end{table}

\subsection{Spectral energy distribution fitting}\label{sec:sedfit}
\begin{figure}
    \centering
    \includegraphics[width=1\columnwidth]{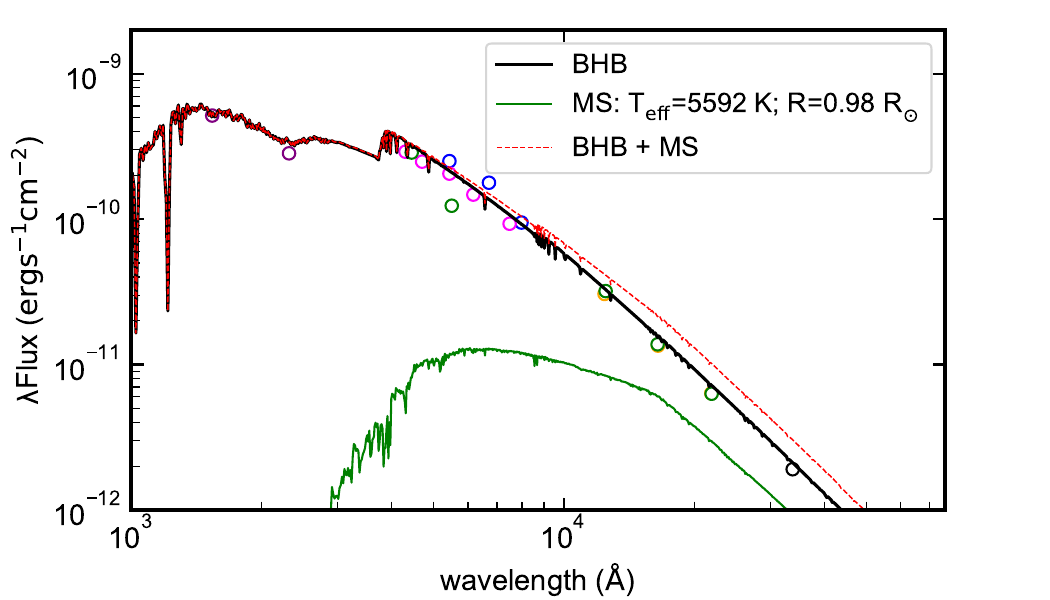}
    \caption{
    The spectral energy distribution, the open dots represent the fluxes from the GALEX $FUV$- and $NUV$-bands (purple), {\it Gaia} EDR3 $G_{\rm BP}$-, $G$- and $G_{\rm RP}$-bands (blue), APASS $B$-, $V$-, $G$-, $R$-, $I$-bands (magenta), JOHNSON $B$-, $V$-, $J$-, $H$-, $K$-bands (green), 2MASS $J$-, $H$- and $K_{\rm s}$-bands (orange), and ALLWISE $W_1$-, $W_2$-bands (black); the black curve shows the best-fitting `kurucz' spectrum of \thestar{}; the red line shows a spectrum of main sequence star with $T_{\rm eff} = 5592$ K, star radius $R = 0.98$ $R_{\odot}$, and without extinction, which is the faintest model of the main sequence star with a mass around 1 $M_{\odot}$ and $0.5 \le Z\le 1$ $Z_{\odot}$ from PARSEC stellar model. The dashed line shows the total SED of BHB and a main-sequence star.     }
    \label{fig:sedfit}
\end{figure}

We used the spectral energy distribution (SED) fitting method to determine the radius, luminosity, extinction, and mass of \thestar{}.
A Python-based program \href{https://speedyfit.readthedocs.io/en/stable/index.html}{\tt SPEEDYFIT} was adopted to perform the fitting procedure. 
During the fitting, the photometric fluxes and their uncertainties from GALEX \citep{GALEX_Bianchi2011Ap&SS.335..161B}, {\it Gaia} EDR3 \citep{GaiaEDR32020yCat.1350....0G}, APASS \citep{APASS_Henden2015AAS...22533616H}, JOHNSON \citep{JOHNSONBV_Hog2000A&A...355L..27H, JOHNSONJHK_IJspeert2021A&A...652A.120I}, 2MASS \citep{2MASS_Cutri2003yCat.2246....0C}, and ALLWISE \citep{ALLWISE_Cutri2012yCat.2311....0C} and parallax measurements from {\it Gaia} DR3, were applied. The parallax was corrected with a zero-point of -0.032 mas \citep{Lindegren2021A&A}. The effective temperature (15524 K) and the surface gravity (4.10 ) were fixed at the values derived from the DBSP spectra.
We used the synthetic spectral model grid `kurucz', which is equipped by {\tt SPEEDYFIT} to fit SED of \thestar.  We derived a radius $0.87\pm0.03$ $R_{\odot}$ and extinction $E(B-V) = 0.073\pm 0.001$ mag. The best-fit SED by a single star agrees very well with the observed multi-band photometry, as shown by the black line in Fig.~\ref{fig:sedfit}. Assuming that the companion object is a main sequence star with a mass of 1 $M_\odot$, we searched for the faintest point ($T_{\rm eff} \sim 5592$ K, $R \sim 0.98$ $R_\odot$) among the isochrones provided by the PAdova and TRieste Stellar Evolution Code (PARSEC) stellar model \citep{Bressan2012MNRAS, Chenyang2014MNRAS} in the intervals of $0.5 \le Z\le 1.7$ $Z_{\odot}$, $1 \le m \le 2$ $M_{\odot}$. If the companion star were visible and had a mass greater than 1 $M_\odot$, there would be an excess of flux in the infrared wavelength range.
This result suggests that the unseen companion's contribution to the observed SED is negligible. So, we obtained a radius of $0.87\pm0.03$ $R_{\odot}$ for the visible companion, and its mass and luminosity could be calculated and listed in Table~\ref{tab:binary}. 

\begin{table*}
\centering
\begin{tabular}{c c c c c } 
\hline
Parameter & Spectroscopic & SED & Light curve & Adopted value\\
\hline
     R.A. (J2000) & --&-- &-- & 12:30:14.92  \\
     Decl. (J2000) & -- & --&-- & +46:37:19.88 \\
     $G$ (mag) & -- & --&-- &12.42960(78)  \\
     $G_{\rm BP}$ (mag) & -- & --&-- &12.3551(20)  \\
     $G_{\rm RP}$ (mag) & -- & --&-- &12.5737(21)  \\
     $\varpi$ (mas) & -- & -- &-- &$ 0.867-(-0.032) \pm0.029$ \\
     $d$ (pc)  & -- & --&-- &$1117^{+37}_{-35}$\\ 
     $\mu_{\alpha}$ (mas\ yr$^{-1}$) & -- & --&-- & $-8.369\pm0.015$ \\
     $\mu_{\delta}$ (mas\ yr$^{-1}$) &  -- & --&-- &$-3.528\pm0.023$ \\
     RUWE &   -- & --&-- & 0.8372 \\
    $T_{\rm eff}^{\rm BHB}$ (K) & $15524\pm307$ & fixed & -- &$15524\pm307$ \\
	$\log g_{\rm BHB}$ (cgs) & $4.10\pm0.03$ & fixed &$4.109_{-0.031}^{+0.029}$ & $4.10\pm0.03$ \\
	$[{\rm Fe/H}]$ (dex) & $-0.17\pm0.20$ & --& -- & $-0.17\pm0.20$ \\
	$M_{\rm BHB}$ ($M_\odot$) & -- &$0.347_{-0.031}^{+0.035}$ & $0.352^{+0.032}_{-0.036}$ & $0.352^{+0.032}_{-0.036}$ \\
    $R_{\rm BHB}$ ($R_\odot$) & -- &$0.87\pm0.03$ & $0.868_{-0.030}^{+0.029}$ & $0.868_{-0.030}^{+0.029}$ \\
	$L_{\rm BHB}$ ($L_\odot$) & -- &$39.8\pm4.1$ & $39.3_{-2.7}^{+2.8}$ & $39.8\pm4.1$ \\
	$M_{\rm comp}$ ($M_\odot$) & -- &-- & $1.26_{-0.14}^{+0.17}$ & $1.26_{-0.14}^{+0.17}$\\
	$T_0$ (BJD-2458900)   & $0.8814(24)$ & -- & 0.8847(29) & 0.8847(29) \\
	$P$ (day)     & fixed &  -- & fixed & 0.8262782(110) \\
	$K_{\rm BHB}$ (\kms) &$188.6_{-3.8}^{+3.9}$ & -- &  $188.0_{-4.7}^{+4.8}$ & $188.6_{-3.9}^{+3.9}$ \\
    $f_{\rm comp}$ ($M_\odot$) &$0.574_{-0.035}^{+0.036}$ &-- & -- & $0.574_{-0.035}^{+0.036}$\\
    $V_{\rm \gamma}$ (\kms)  &$-4.9\pm2.7$ &-- & -- & $-4.9\pm2.7$\\
	$q$           & -- & -- & $3.54_{-0.54}^{+0.67}$ & $3.54_{-0.54}^{+0.67}$ \\
    $e$           & fixed & -- & fixed & 0 \\
    $i$ (\degree) & --& --   & $65.3_{-6.1}^{+8.8}$ & $65.3_{-6.1}^{+8.8}$ \\
    $v_{\rm rot}\sin i$ (\kms) & $48.3\pm2$& --  & $48.2_{-2.6}^{+2.8}$ & $48.3\pm2$ \\
    $a$ ($R_\odot$) & --& -- & $4.34_{-0.13}^{+0.14}$ & $4.34_{-0.13}^{+0.14}$ \\
	$R^{\rm Roche-lobe}_{\rm BHB}$ ($R_{\odot}$) & --& -- &  $1.199_{-0.038}^{+0.041}$ & $1.199_{-0.038}^{+0.041}$\\
	$R^{\rm Roche-lobe}_{\rm comp}$ ($R_{\odot}$) & --& --   & $2.13_{-0.12}^{+0.13}$ & $2.13_{-0.12}^{+0.13}$ \\
   \hline
\end{tabular}
   \caption{\textbf{Astrometric, stellar, and orbital parameters of \thestar{}}.  Astrometric parameters are from {\it Gaia} DR3 \citep{Gaiadr32023A&A...674A...1G}, with a zero-point (-0.032 mas) correction applied to the parallax \citep{Lindegren2021A&A}. For stellar and orbital parameters, we show the values obtained from spectroscopic analyses, SED fitting, and the TESS LC fitting by using {\tt LCURVE} \citep{Copperwheat2010MNRAS.402.1824C}, if derived independently, as well as the adopted values which result from the LC solutions. The values quoted are the median, and the uncertainties give the confidence interval 68\%. Quantities shown are the right ascension RA, declination DEC, {\it Gaia} magnitude $G$, {\it Gaia} BP-band magnitude $G_{\rm BP}$, {\it Gaia} RP-band magnitude $G_{\rm RP}$, parallax $\varpi$, proper motions in the right ascension $\mu_\alpha$ and declination $\mu_\delta$ directions, the Renormalised Unit Weight Error RUWE, the effective temperature $T_{\rm eff}^{\rm BHB}$, surface gravity $\log~g_{\rm BHB}$, iron abundance $[{\rm Fe/H}]$, 
   mass $M_{\rm BHB}$, radius $R_{\rm BHB}$, and luminosity $L_{\rm BHB}$, the companion star mass $M_{\rm comp}$,the zero-point of the ephemeris $T_0$, the orbital period $P$,  the radial-velocity semi-amplitude $K_{\rm BHB}$, the binary mass function of unseen companion $f_{\rm comp}$,
   the systemic radial velocity of the binary $V_\gamma$, the mass ratio $q$, 
   the eccentricity $e$, 
   the orbital inclination $i$, the projected rotational velocity $v_{\rm rot}\sin i$, the orbital distance $a$, the effective Roche-lobe radii of the BHB star $R^{\rm Roche-lobe}_{\rm BHB}$ and the companion star $R^{\rm Roche-lobe}_{\rm comp}$. 
   }
   \label{tab:binary}
\end{table*}

\subsection{Radial velocity curve fitting}\label{sec:rvfit}

We model the radial velocity measurements from P200, Keck-I, and X216 using an MCMC framework implemented with {\tt EMCEE}
\citep{emcee2013ascl.soft03002F}. The likelihood function takes the form
\begin{equation}\label{eq:proprv}
   \begin{split}
    \ln[\mathcal{L}(\theta_v \mid v)] &\propto \ln [\mathcal{L}(v \mid \theta_v)] \\
    &= -\frac{1}{2}\sum_{i=1}^{N}\frac{[v_{i}-v(\theta_v, t_i)]^2}{\sigma^2_{v_i}+s^2}-\frac{1}{2}\sum_{i=1}^{N}\ln[2\pi(\sigma_{v_i}^2+s^2)],
  \end{split}
\end{equation}
where $v_{i}$ and $\sigma_{v_{i}}$ are the radial velocities and their measurement errors, respectively, and $\theta_v = $($P$, $T_0$, $K_{\rm BHB}$, $V_\gamma$, $\sqrt{e}\cos\omega$, $\sqrt{e}\sin\omega$) are the orbital parameters.
$K_{\rm BHB}$, $e$, $\omega$, and $V_\gamma$ are the radial velocity semi-amplitude of the BHB, eccentricity, longitude of periastron, and systematic radial velocity, respectively. 

We fix the orbital period at $P = 0.8262782$~d from the TESS light curve. Given that the phase separation between adjacent maxima is approximately 0.5, we also adopt a circular orbit ($e = 0$). The model radial velocity $v(\theta_v, t_i)$ is computed using the
{\tt rv\_drive} function in the {\tt radvel} package
\citep{RadVel_Fulton2018}.

An additional jitter term $s$ is included in quadrature with the measurement uncertainties to account for unmodeled effects, such as instrumental systematics, wavelength calibration errors, and possible model imperfections. The best-fitting model is shown as the black curve
in Fig.~\ref{fig:lcrvfit}(b). We obtain $K_{\rm BHB} =
188.4_{-3.8}^{+3.9}$~\kms{} and $V_\gamma = -4.9\pm2.7$~\kms{}. The corresponding binary mass function is
\begin{equation}\label{eq:massfunc}
    f_{\rm comp} = \frac{M^3_{\rm comp}\sin^3i}{(M_{\rm BHB} + M_{\rm comp})^2}
    = \frac{PK^3_{\rm BHB}}{2\pi G}
    = 0.574^{+0.036}_{-0.035}\, M_{\odot},
\end{equation}
where $G$ is the gravitational constant. The fitting results are also summarized in Table~\ref{tab:binary}.

\subsection{Light curve fitting}\label{sec:lcfit}
Due to the lack of infrared excess in the SED fit, we assumed that the ellipsoidal variation was generated by the light from the BHB surface, which was caused by the strong tidal force of its companion. The semi-amplitude of the ellipsoidal light curve can be presented as \citep{Morris1993ApJ...419..344M, Zucker2007ApJ...670.1326Z}
\begin{equation}\label{eq:deltaF}
    \frac{\Delta F_{\rm BHB}}{F_{\rm BHB}} \backsimeq 0.15 \frac{(15+\epsilon_{\rm BHB})(1+\tau_{\rm BHB})}{3-\epsilon_{\rm BHB}}\left(\frac{R_{\rm BHB}}{a}\right)^3 \frac{M_{\rm comp}}{M_{\rm BHB}} \sin^2i
\end{equation}
where $\epsilon_{\rm BHB}$ is the linear limb-darkening coefficient, $\tau_{\rm BHB}$ is the gravity darkening exponent, $R_{\rm BHB}$ is the radius of the BHB companion, $i$ is the inclination, $a$ is orbital distance,
\begin{equation}\label{eq:a}
a^3 = G(M_{\rm comp} + M_{\rm BHB}) \left(\frac{P}{2\pi}\right)^2.
\end{equation}
$R_{\rm BHB}$ has been measured by SED fit. The atmosphere parameters of BHB have been estimated from DBSP spectra, so  $\epsilon_{\rm BHB}$, $\tau_{\rm BHB}$ and $m_{\rm BHB}$ can be obtained. Therefore, the semi-amplitude of the ellipsoidal light curve is a function of $m_{\rm comp}$ and $i$. Combining binary mass function Eq.~(\ref{eq:massfunc}),  we can place constraints on the mass of the unseen companion $m_{\rm comp}$ and $i$ from the semi-amplitude of the ellipsoidal light curve.

Rather than simply using Eq.~(\ref{eq:deltaF}), we model the light curve in a more complete way using {\tt LCURVE} \citep{Copperwheat2010MNRAS.402.1824C}, and perform an MCMC approach to fit the TESS light curve. {\tt LCURVE} utilizes a grid of points to model the shapes of the two stars, which are influenced by the Roche potential. The flux emitted from each point on the grid is calculated from a blackbody model with an estimated temperature at the reference wavelength of 7697.6 \AA{}, corresponding to the TESS bandpass\footnote{\url{http://svo2.cab.inta-csic.es/theory/fps/}}.

We assume a circular orbit ($e=0$) and fix the orbital period at 0.8262782 days. Before conducting the light-curve fitting, we phase-fold it using the ephemeris from the radial-velocity solution, bin it with a phase interval of 0.01, and remove $3\sigma$ outliers in each bin.  

Test runs indicate that the companion’s temperature and radius cannot be constrained due to its negligible flux contribution. We therefore fix these parameters to $T_{\rm eff} = 10{,}000$ K and $R = 0.01,R_\odot$, consistent with a white dwarf companion.

For the BHB star, we fix $T_{\rm eff} = 15{,}524$ K and adopt a Doppler beaming factor of 1.6, calculated from rotationally broadened synthetic spectra based on the BSTAR2006 grid following Equation~(4) of \citet{Claret2020A&A...641A.157C}. The limb- and gravity-darkening coefficients are interpolated from \citet{Claret2017A&A...600A..30C}. These coefficients are set to zero for the companion due to its negligible contribution.

We adopt Gaussian priors on the spectroscopically and SED-derived parameters, with $\log g_{\rm BHB} = 4.10 \pm 0.03$ (cgs), $R_{\rm BHB} = 0.87 \pm 0.03$ $R_\odot$, and $K_{\rm BHB} = 188.6 \pm 3.9$ \kms{}. A standard $\sin i$ prior is assumed for the inclination, and a uniform prior is adopted for $T_0$ within a narrow range around the radial velocity solution. The remaining parameters are then constrained through MCMC fitting of the light curve. Ignoring the flux contribution from the companion, we obtain an orbital inclination of $65.3_{-6.1}^{+8.8}\degree$ which then gives the companion mass to be  $1.26^{+0.17}_{-0.14}$ $M_{\odot}$. The fitting results are also listed in Table~\ref{tab:binary}. 

According to dynamical tidal theory, synchronization in an sdB star proceeds from the outer envelope toward the inner core (see Section 5.3 of \citealt{Ma2024ApJ...975....1M}). Assuming that the surface of the BHB component is synchronized, the corresponding projected rotational velocity is $ 48.2^{+2.8}_{-2.6}$~\kms, in agreement with the spectroscopic measurement ($48.3\pm2$~\kms). This consistency indicates that the surface of the BHB has likely reached synchronous rotation. However, the current light curve does not provide sufficient information to constrain the internal rotation period, and therefore, it remains unclear whether the interior has also achieved synchronous rotation.

Considering potential systematic uncertainties in the surface gravity derived from the P200 spectra using the TLUSTY/BSTAR2006 grid \citep{Przybilla2011JPhCS.328a2015P}, we perform a test by inflating the uncertainty in $\log g_{\rm BHB}$ to 0.1 dex and adopting a broader Gaussian prior. The corresponding results are listed in Table~\ref{tab:diff_prior}. We obtain a companion mass of $1.15^{+0.22}_{-0.10}$ $M{\odot}$ and a BHB mass of $0.341^{+0.043}_{-0.049}$ $M{\odot}$. These values are consistent with those derived using the fiducial prior, which yield $1.26^{+0.17}_{-0.14}$ $M{\odot}$ and $0.352^{+0.036}_{-0.032}$ $M{\odot}$, respectively.

\begin{table}
\centering
\begin{tabular}{c c |c|c} 
\hline
\multirow{1}{*}{Prior}
 &$\log g_{\rm BHB}$ & $\mathcal{N}(4.10, 0.03)$ & $\mathcal{N}(4.1, 0.1)$\\
   \hline
\multirow{4}{*}{Posterior}
& $\log g_{\rm BHB}$ (cgs) & $4.109_{-0.031}^{+0.029}$ & $4.097_{-0.063}^{+0.032}$\\
&$i$ (\degree)  & $65.3_{-6.1}^{+8.8}$ & $72.4_{-12.9}^{+12.0}$\\
& $R_{\rm BHB}$ ($R_\odot$) & $0.868_{-0.030}^{+0.029}$ & $0.871_{-0.030}^{+0.029}$\\
& $T_0$ (BJD-2458900) & $0.8847_{-0.0025}^{+0.0029}$ & $0.8847_{-0.0025}^{+0.0028}$\\ 
&$K_{\rm sd}$ (\kms)   &  $188.3_{-4.7}^{+4.8}$ & $188.7_{-3.9}^{+3.8}$\\
\hline
\multirow{7}{*}{Calculation}
&$q$ & $3.61_{-0.54}^{+0.67}$ & $3.30_{-0.45}^{+1.21}$ \\
& $M_{\rm BHB}$ ($M_\odot$) & $0.352_{-0.036}^{-0.032}$ & $0.341_{-0.049}^{+0.043}$\\
&$M\comp$ ($M_\odot$) & $1.26_{-0.14}^{+0.17}$ & $1.15_{-0.10}^{+0.22}$\\
&$M_{\rm all}$ ($M_\odot$) & $1.61_{-0.13}^{+0.16}$ & $1.51_{-0.11}^{+0.18}$\\
& $v \sin i$ (\kms) & $48.2_{-2.6}^{+2.8}$ & $50.4_{-4.7}^{+2.9}$\\
&$L_{\rm BHB}$ ($L_\odot$)&  $39.3_{-2.7}^{+2.8}$ & $39.7_{-2.6}^{+2.7}$\\
&$a$ ($R\odot$)  &  $4.34_{-0.13}^{+0.14}$ & $4.25_{-0.11}^{+0.16}$\\
&$R^{\rm Roche-lobe}_{\rm BHB}$ ($R_{\odot}$) & $1.199_{-0.038}^{+0.041}$ & $1.186_{-0.061}^{+0.050}$\\
&$R^{\rm Roche-lobe}\comp$ ($R_{\odot}$)  & $2.13_{-0.12}^{+0.13}$ & $2.047_{-0.081}^{+0.197}$\\

\hline
\end{tabular}
   \caption{The results of fitting the light curve using different prior distributions of the surface gravity of the BHB component.
   Quoted values are the median, and uncertainties denote the 68\% confidence interval. 
   The quantities shown are the surface gravity $\log~g_{\rm BHB}$, the orbital inclination $i$, the BHB component radius $R_{\rm BHB}$, the superior conjunction time $T_0$, the radial velocity semi-amplitude $K_{\rm BHB}$, the mass ratio $q$, the BHB component mass $M_{\rm BHB}$, the companion star mass $M_{\rm comp}$, the total mass of the binary $M_{\rm all}$, the projected rotation velocity $v\sin i$ (assuming synchronous rotation), the BHB component luminosity $L_{\rm BHB}$, the orbital separation $a$, the effective Roche-lobe radii of the BHB component $R^{\rm Roch-lobe}$ and the companion star $R^{\rm Roche-lobe}\comp$.
   }
   \label{tab:diff_prior}
\end{table}

\section{Nature and Evolution of the Feige 64}\label{sec:theory}

\subsection{The Galactic Kinematics}\label{sec:kinematice}
Using the {\it Gaia} astrometry and systemic velocity, we derive the 3D Galactic velocity of \thestar. In our calculations, we assumed that the Sun is located on the Galactic disk at $z=27$ pc (ref.\cite{Chenbing2001ApJ...553..184C}), at a distance of 8.27 kpc from the Galactic center, that its local circular velocity is 238 \kms{} \citep{Schonrich2012MNRAS.427..274S}, and that its peculiar velocities respected to the Local Standard of Rest is 
$(U_{\odot}, V_{\odot}, W_{\odot}) = (9.58, 10.52, 7.01)$ \kms{} \citep{Tianhaijun2015ApJ...809..145T}. The distances of \thestar{} to the Galactic Center and the Galactic plane are $d_{\rm GC} = 8.670\pm0.013$ kpc and $d_{z} = 1.074\pm0.034$ kpc, respectively. The resulting cylindrical coordinate velocity is $(V_R, V_{\phi}, V_Z) = (25.4\pm1.3, 209.9\pm1.4, 4.9\pm2.5)$ \kms.
The Galactic orbits (see Fig.~\ref{fig:space_trajectory}) for \thestar{} were calculated
in the Galactic potential Model I of \cite{Irrgang2013A&A...549A.137I} with a Bulirsch-Stoer integrator. The orbits show the eccentricity of Galactic kinematics $e_{\rm GC} \sim 0.14$. The Galactic kinematics and eccentricity indicate that \thestar{} is from the thin disk population.

Following the approach of \cite{Chendichang2021ApJ...909..115C}, the total velocity dispersion of \thestar{} in the Galaxy is estimated to be $\sim57.2$ \kms{}, and the kinematic age of $\sim 6.5$ Gyrs is calculated following the age-velocity dispersion relation. This age implies that the visible companion could not have evolved in isolation from an initial mass below 0.8 $M_\odot$, as the hydrogen-burning timescale for such low-mass stars exceeds the age of the Universe.

\subsection{The nature of the visible component of Feige 64}\label{sec:visiblenature}

Given the location of Feige 64 in the HR diagram (Fig. ~\ref{fig:cmd}) and the mass of the visible component, the visible component appears to be like a blue horizontal branch (BHB) star, i.e., a helium-core-burning star with a thick envelope.

Following the method of \cite{Arancibia-Rojas2024MNRAS.52711184A}, we construct a series of BHBs (or sdBs) models with different helium cores and envelope masses using a stellar evolution code \textsc{mesa} \citep{pbdh+11,pcab+13,pmsb+15,psbb+18,pssg+19}. 
We first create a series of stars at the tip of the red giant branch (T-RGBs) by using main-sequence stars with initial masses ranging from 1.85 $M_\odot$ to 2.10 $M_\odot$ at a step of 0.05 $M_\odot$. We set the initial metallicity to be 0.018, as the [Fe/H] of \thestar\, is consistent with solar iron abundance (see Table~\ref{tab:abundanceVt=8}), and \thestar\, is a member of the thin disk population. Their helium core masses ($M_{\rm He core}$) are 0.3628, 0.3441, 0.3372, 0.3338, 0.3320, and 0.3317 $M_\odot$, respectively.

We do not consider main-sequence progenitors with initial masses below 1.85 $M_\odot$. For lower-mass stars ($M_{\rm ZAMS}\lesssim2,M_\odot$), the helium core becomes increasingly electron-degenerate during RGB evolution, causing helium ignition to occur at a larger core mass near the tip of the RGB (see \citealt{Hanzhanwen2002MNRAS.336..449H, Arancibia-Rojas2024MNRAS.52711184A} and references therein). The lowest-mass progenitor considered here (1.85 $M_\odot$) already produces a helium core mass of 0.3628 $M_\odot$, which exceeds the estimated mass of the BHB component of \thestar{} ($\sim0.352\,M_\odot$), and lower-mass progenitors would result in even larger helium cores.

We then artificially strip the envelope of the T-RGBs with the "relax\_mass" routine in \textsc{MESA}, constructing a grid of BHB (or sdB) models with masses ($M_{\rm model}$) ranging from 0.34 to 0.39 $M_\odot$ at a step of 0.0005 $M_\odot$, subject to $M_{\rm model} > M_{\rm He core} + 0.0005 M_{\odot}$. An overshooting parameter of 0.016 is adopted, following standard prescriptions for low- and intermediate-mass stars \citep{Herwig2000A&A...360..952H, Claret2017ApJ...849...18C}.

\begin{figure*}
\centering
\includegraphics[scale=0.68, angle=0]{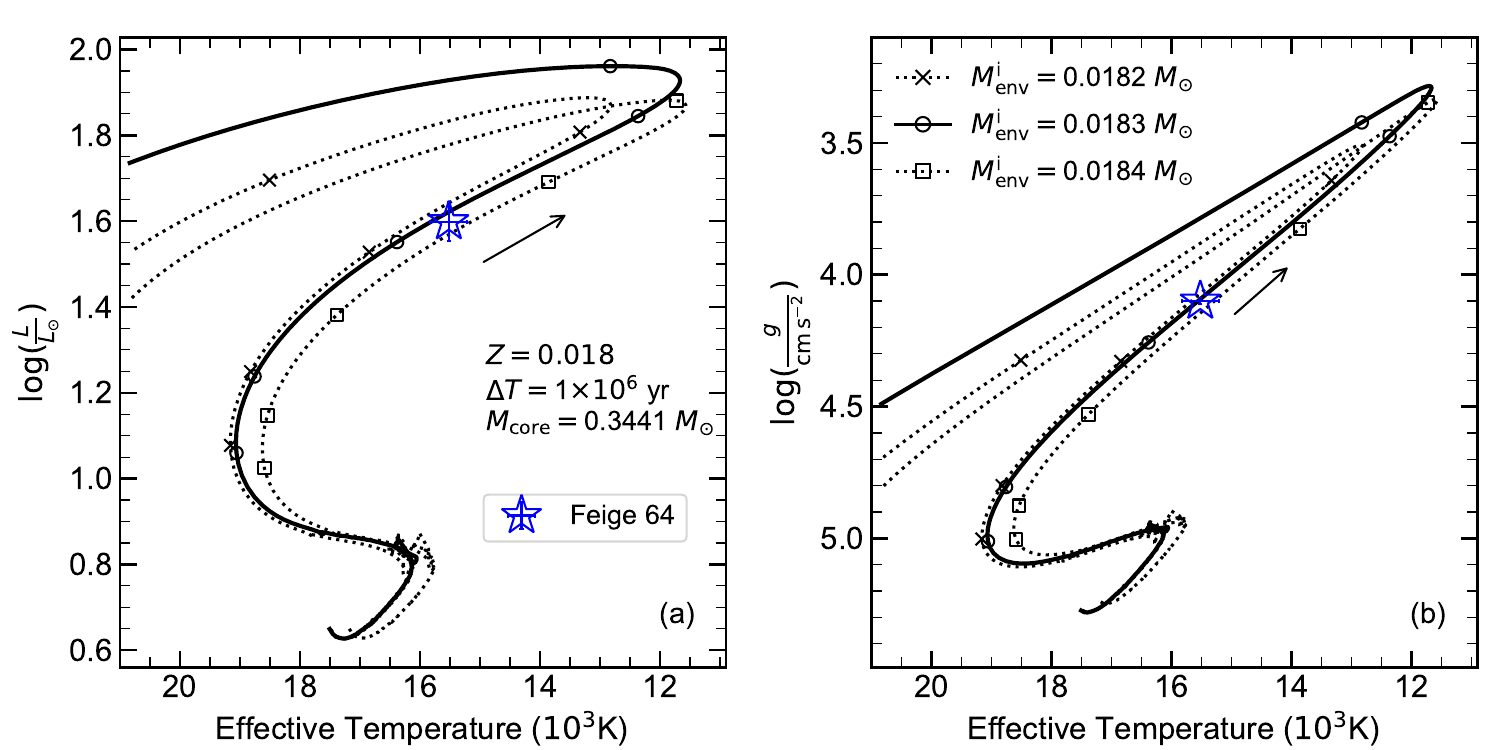}
\caption{Evolution tracks are shown in the Hertzsprung-Russell diagram (a) and the Kiel diagram (b). These models, constructed by artificially stripping the envelope of a 1.9\;$M_\odot$ progenitor around its tip of red giant branch, have an initial helium core mass of $0.3441\;M_\odot$. The curved lines marked by crosses, circles, and squares represent the evolutionary tracks of models with initial envelope masses of 0.0182, 0.0183, and 0.0184 $M_\odot$, respectively. Markers are placed at intervals of one million years, starting from the point where $\log g = 5.0$ (cgs).  The blue star marks the location of the observed \thestar
. The arrows indicate the directions of evolution. }
\label{fig:hr_kiel}
\end{figure*}

We find that models with masses in the range $\sim 0.345$–$0.365~M_\odot$ can broadly reproduce the observed position of \thestar\ in both the HR and Kiel diagrams. However, for progenitor masses $\gtrsim 1.95~M_\odot$, the binding energy of the envelope exceeds the available orbital energy, making successful common-envelope ejection unlikely. We therefore refine our grid using a progenitor mass of 1.9~$M_\odot$, exploring models with total masses between 0.35 and 0.365~$M_\odot$ at a finer resolution (step size 0.0001~$M_\odot$). The best-fitting model has a helium-core mass of $0.3441~M_\odot$ and an envelope mass of $0.0183~M_\odot$, and reproduces the observed properties in both diagrams (Fig.~\ref{fig:hr_kiel}). The model persists in the BHB-like phase for approximately 2~Myr (as shown in Fig.~\ref{fig:hr_kiel}), during which $\log g$ decreases from $\sim 4.5$ to $\sim 3.4$ (cgs).

\begin{figure}
    \centering
    \includegraphics[width=1\columnwidth]{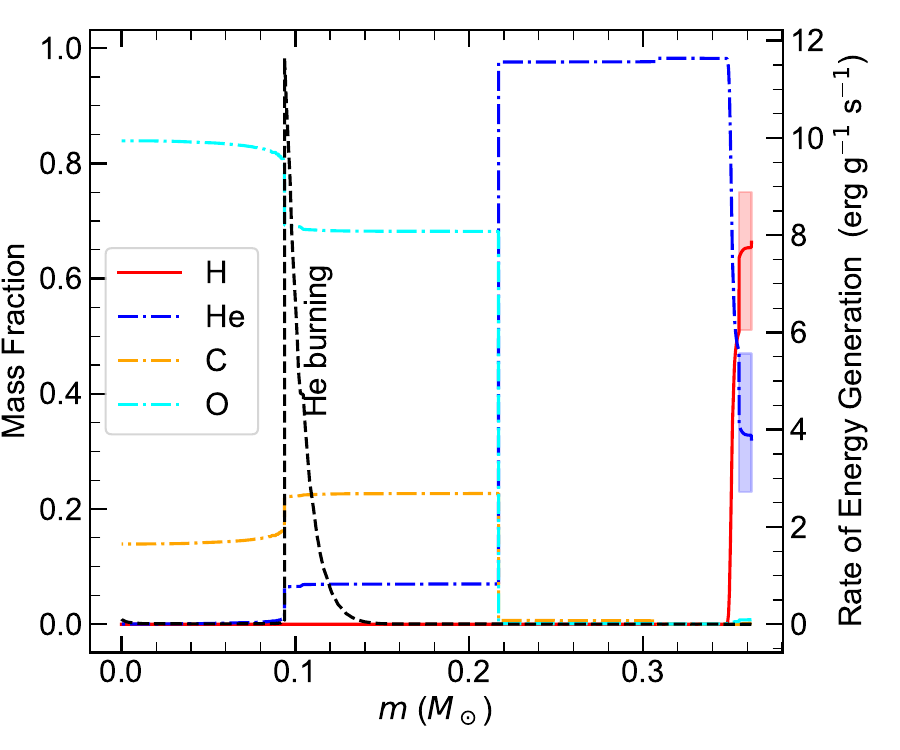}
    \caption{The profile of the best-fitting model. \textbf{Left axis} is for the element mass fraction. The red solid, blue dashed-dotted, red dashed-dotted, and cyan dashed-dotted lines represent the mass fractions of hydrogen, helium, carbon, and oxygen, respectively. The blue and red shaded areas indicate the 1$\sigma$ uncertainty regions for the surface Helium and Hydrogen mass fractions, respectively, as derived from the Keck-I spectrum. \textbf{Right axis}, the black dashed line represents the profile of the rate of energy generation for helium burning.}
    \label{fig:profile}
\end{figure}

The internal structure of this model at the current evolutionary stage of \thestar\, is shown in Fig.~\ref{fig:profile}. Helium is nearly exhausted in the core, with a central mass fraction below $10^{-3}$, while helium-shell burning remains active (black dashed line). The core is primarily composed of carbon and oxygen, and the luminosity is dominated by hydrogen-shell burning. The predicted surface hydrogen and helium abundances are also consistent with those derived from the Keck-I spectrum.

From these models, we infer a helium-core mass of $\sim 0.344~M_\odot$ and an envelope mass of $\sim 0.018~M_\odot$ for the BHB star. Its progenitor is likely a $\sim 1.9~M_\odot$ main-sequence star.

\subsection{Is the unseen companion star a normal star or compact object?}\label{sec:compactnature}

Adopting the BHB mass derived from the SED and spectroscopic analysis, together with the radial velocity semi-amplitude and orbital period, we compute the projected semi-major axis ($a\cos i$) as a function of inclination. 
As shown in the TESS light curve (Fig.~\ref{fig:lcrvfit}), no eclipse is detected, implying that the orbital configuration must satisfy $a\cos i > (R_{\rm BHB} + R_{\rm comp})$. Assuming the unseen companion is a main-sequence star, its minimum radius at a given mass (and hence inclination) can be estimated using PARSEC models. Under this assumption, eclipses would occur for $i \gtrsim 62.6^\circ$. The absence of eclipses therefore requires $i \lesssim 62.6^\circ$, which in turn implies $M_{\rm comp} \gtrsim 1.2,M_\odot$ from the mass function. However, such a massive main-sequence companion would produce a clear infrared excess in the SED (Section~\ref{sec:sedfit}), which is not observed. This rules out a main-sequence companion.


We next consider the possibility of a hot subdwarf companion. Adopting typical parameters ($M \sim 0.45,M_\odot$, $\log g \sim 5.5$, $T_{\rm eff} \sim 25{,}000$ K), its radius would be $\sim 0.2,R_\odot$ (refs. \cite{Hanzhanwen2003MNRAS.341..669H, Heber2016PASP..128h2001H}). In this case, the flux contribution in the optical band would exceed $\sim 7\%$, which should produce detectable spectral features. Even under an extreme assumption of a 10\% flux contribution, the inferred minimum radius of the visible companion can be estimated as $\sim 0.82$ $R_\odot$, corresponding to a minimum mass of $\sim 0.3$ $M_\odot$. For an inclination of 90\degree, the binary mass function further implies a minimum companion mass of $\sim 0.97,M_\odot$. Such a massive hot subdwarf would be significantly brighter than the visible component, which is inconsistent with the observations.

Taken together, these arguments strongly suggest that the unseen companion is a compact object.
Assuming the companion is a neutron star, we estimate the expected X-ray emission from wind accretion. Using the wind prescription of \citet{Vink2001A&A...369..574V}, the BHB mass-loss rate is $\sim 3.7\times10^{-12},M_\odot,{\rm yr^{-1}}$. If 1\% of this wind is accreted, the expected X-ray flux in the 0.3–2.3 keV band is $\sim 2.9\times10^{-13}\,{\rm erg\cdot cm^{-2} \cdot s^{-1}}$. However, no counterpart is detected in SRG/eROSITA data \citep{Sunyaev2021A&A...656A.132S, Predehl2021A&A...647A...1P}, which places a $3\sigma$ upper limit of $1.5\times10^{-14}\, {\rm erg\cdot cm^{-2} \cdot s^{-1}}$. Combined with the non-detection of radio pulsations, this disfavors a neutron star companion. We therefore conclude that the unseen companion is most likely a WD, although a neutron star cannot be entirely ruled out.

\subsection{Formation and evolution of the Feige 64 binary}\label{sec:evolution}

The observed $T_{\rm eff}$, $\log g$, near-solar metallicity, and derived mass demonstrate that the visible star is a stripped remnant currently undergoing helium-shell burning. An interpretation as a proto–extremely low-mass WD (ELM WD) can be ruled out, since typical ELM WDs have $M \approx 0.27\,M_\odot$ \citep{Althaus2013A&A...557A..19A} and fail to reproduce the observed mass–period relation for stable mass transfer. Moreover, forming a proto-ELM WD through CEE would require an unrealistically high efficiency, $\alpha_{\rm CE} > 10$ \citep{lizhenwei2019ApJ...871..148L}, rendering this channel implausible for \thestar{}.

The short orbital period of the \thestar{} binary indicates that it cannot be produced from a stable mass transfer channel, but should be produced from a common-envelope (CE) channel \citep{Hanzhanwen2002MNRAS.336..449H}. In the CE channel, a binary consisting of a white dwarf and a red giant star undergoes unstable mass transfer when the red giant star is near the tip of the red giant branch, after which the binary enters the common-envelope phase. After the ejection of the CE, the binary system will evolve into the \thestar{} binary. 

To investigate this process, we adopt the energy-budget prescription for the evolution of CE phase,
\begin{equation}
\begin{split}
\int_{M_{\mathrm{core}}}^{M_2} & \left(-\frac{G M(r)}{r}+U\right) \mathrm{d} m \\
&= \alpha_{\mathrm{CE}} \left(-\frac{G M_{\mathrm{core}} M_{\rm comp}}{2 a_{\mathrm{f}}} + \frac{G(M_{\mathrm{core}}+M_{\mathrm{env}}) M_{\rm comp}}{2 a_{\mathrm{i}}}\right),
\end{split}
\label{eq:ce}
\end{equation}
where $M_{2} = 1.90\;M_{\odot}$, $M_{\rm core} = 0.362\;M_{\odot}$, $M_{\rm comp} = 1.26\;M_{\odot}$ and $M_{\rm env} = M_{2}-M_{\rm core} = 1.538\;M_{\odot}$. Here $U$ is the internal energy and $M(r)$ is the mass of the star as a function of the radius $r$. $a_{\rm i}$ and $a_{\rm f}$ are the binary separation at the onset and end of CEE, respectively. $\alpha_{\rm CE}$ is the ejection efficiency.
The binding energy, i.e, the left part of the equation, can be computed from the stellar model with a mass of $1.90\;M_{\odot}$ around the tip of the red giant branch.
The binary separation $a_{\rm i}$ is determined by the condition that the radius of the red giant star equals its Roche lobe radius.
$a_{\rm f}$ is the current binary separation. Then we obtain $\alpha_{\rm CE} = 0.90$, which is less than 1. This means that the \thestar{} system can be produced from CE channel. 

As discussed in Section~\ref{sec:visiblenature}, the post-CE remnant is expected to remain in the BHB phase for $\sim 2$ Myr, retaining a hydrogen-rich envelope of $\sim 0.018\, M_\odot$, which is more massive than previously recognized. In the low-mass regime, post-CE remnants were generally thought to preserve no more than $0.01\, M_\odot$ of hydrogen envelope, appearing observationally as hot subdwarfs \citep{Hanzhanwen2002MNRAS.336..449H, Gehongwei2022ApJ...933..137G} or helium-core white dwarfs \citep{Althaus2025A&A...699A.280A}.

In contrast, our results suggest that \thestar{} originated from a $1.9\, M_\odot$ progenitor, which ignites helium under a non-degenerate condition and has a rather extended transition region (of order 0.016 $M_\odot$) between the hydrogen-exhausted core and the envelope (see also \citealt{Justham2011MNRAS.410..984J}). This implies that post-CE systems can retain a thicker hydrogen shell than previously assumed, providing a valuable new constraint on the physics of CE evolution.

\begin{figure}
    \centering
    \includegraphics[width=1\columnwidth]{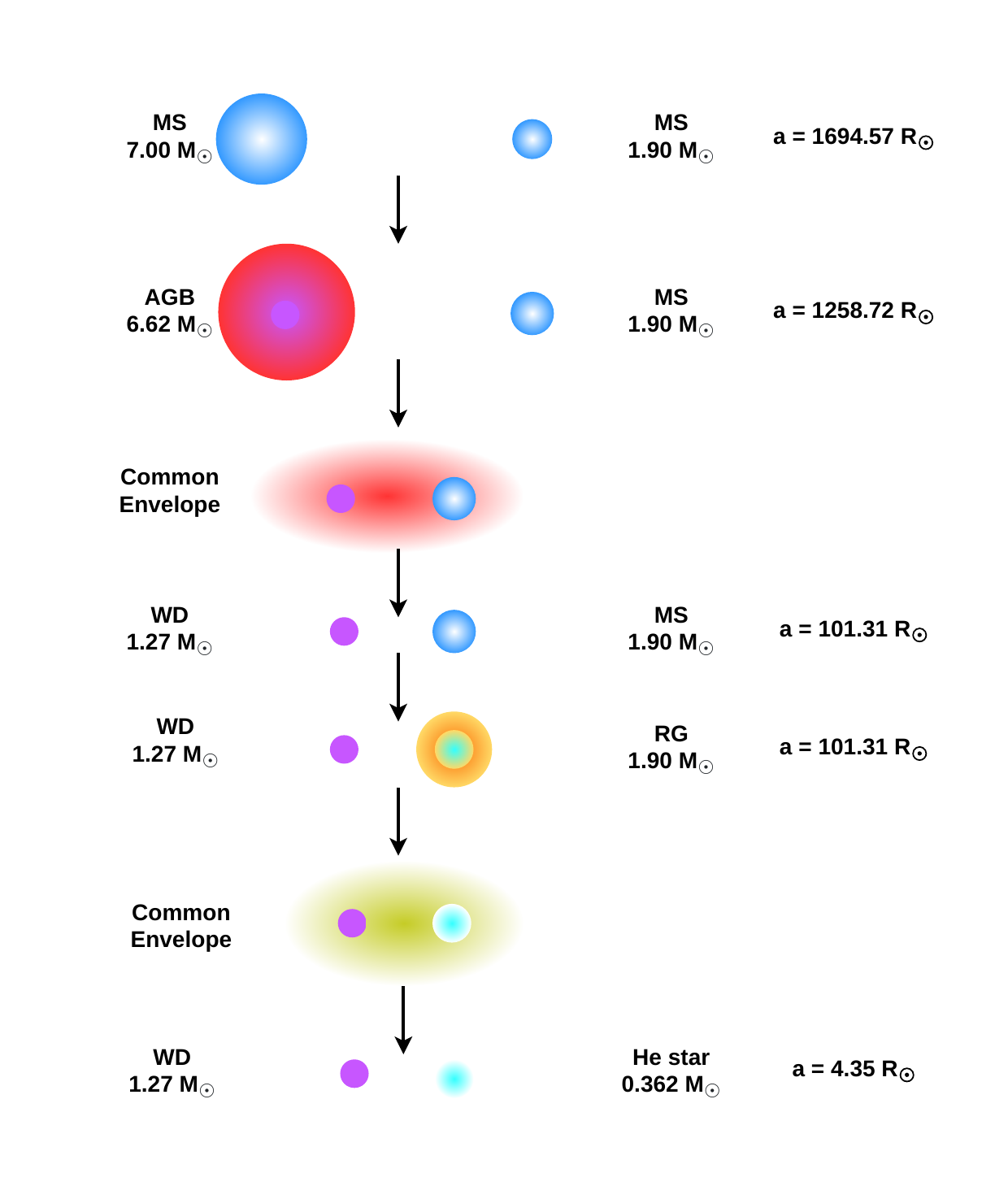}
    \caption{The evolution cartoon of \thestar. The observed properties can be produced after twice common envelope ejections (CEEs). For the second CEE, the ejection efficiency $\alpha_{\rm CE}$ is set to 0.9.}
    \label{fig:evol_bse}
\end{figure}

The mass of the white dwarf suggests a progenitor with an initial mass of approximately
$6.0-8.0\;M_{\odot}$. Given that the wind accretion is very inefficient and little mass is accreted during CE evolution, the initial main-sequence mass of the secondary star is expected around $1.90\;M_{\odot}$. The initial orbital period is assumed to be $2000-3000\;$days. In such a system, the massive primary evolves more rapidly to the asymptotic giant branch (AGB) phase and initiates dynamically unstable mass transfer, leading to the first CE phase. After the successful CEE, the core of the AGB star evolves into a WD, producing a binary consisting of a WD and a main-sequence star. As the secondary evolves to the tip of the red giant branch, a second phase of unstable mass transfer occurs, triggering another CE episode. Following this second CEE, a binary system consisting of a WD and a helium star with a thick envelope is produced and later evolves into the present-day \thestar{} binary system. A specific example for the formation of \thestar{} is shown in Fig.~\ref{fig:evol_bse}.

For the future evolution, we model the system using \textsc{mesa}. The secondary is expected to expand and undergo a short phase of mass transfer. Once the hydrogen shell is stripped, the system will become a double compact binary. Its orbit will decay due to gravitational-wave radiation. According to the Peters-Mathews formalism \citep{Peters1964PhRv..136.1224P}, the timescale is $\sim 75$ Gyr. Therefore, the system will not evolve into a semi-detached configuration within a Hubble time.


\section{Conclusion and discussion}\label{sec:conclusion}

Combined analysis of Gaia DR3 astrometry and kinematic modeling places \thestar{} in the Galactic thin disk with a kinematic age of $\sim$6.5 Gyr. This age implies that the visible component cannot have evolved in isolation from a progenitor with $M \lesssim 0.8\,M_\odot$, as such stars have main-sequence lifetimes longer than the age of the Universe.

The observed properties ($T_{\rm eff}$, $\log g$, and mass) identify the visible component as a stripped remnant in the helium-shell-burning phase. An interpretation as a proto–ELM WD is disfavored by both the observed mass–period relation and the energetically implausible CE efficiency ($\alpha_{\rm CE} > 10$) required for this formation channel. Instead, our models indicate \thestar{} is a BHB star that retained a relatively thick hydrogen-rich envelope ($\sim 0.018\,M_\odot$) following CEE. This thicker envelope acts as a fossil record of the extended core-envelope transition region in its $1.9\,M_\odot$ progenitor, providing a unique constraint on the physics of envelope stripping.

Notably, the measured rotational velocity ($v_{\rm rot} \sin i \sim 48$~\kms) of Feige 64 marks it as a significant outlier within the broader BHB population. While the majority of BHB stars with $T_{\rm eff} > 11,500$~K exhibit negligible rotation due to radiative levitation (\citealt{Catelan2009Ap&SS.320..261C} and references therein), the rapid rotation of Feige 64 is likely a consequence of its close binary nature, where strong tidal torques in the post-CEE stage have synchronized the envelope rotation with the short orbital period.

Detailed binary population synthesis suggests a double-CE channel for the formation of Feige 64: a $\sim 7\,M_\odot$ primary first produced a WD via a CEE, followed by a second CEE initiated when the $\sim 1.9\,M_\odot$ secondary reached the tip of the RGB. This sequence successfully produced the present-day close binary. \thestar{} stands as a rare, observationally confirmed post-CEE remnant where concurrent helium- and hydrogen-shell burning sustains its BHB-like luminosity. It thus provides a critical evolutionary link for understanding the formation of BHB stars through the common-envelope channel, demonstrating that a subset of metal-rich BHB stars are in fact helium-shell-burning stripped stars in compact binaries rather than canonical core-helium-burning objects.

\section*{Acknowledgements}

We are grateful to M\'{a}rcio Catelan and Aldo Valcarce for insightful discussions on BHB stars. We also thank Dongdong Liu, Chenyuan Wu, Weiwei Zhu, Shuai Zha, and many others for their valuable contributions to this project. We acknowledge the support of the staff of the Lijiang 2.4m telescope and the Xinglong 2.16m telescope.

This work is supported by the National Natural Science Foundation of China (NSFC, Nos. 12288102, 12090040/3, 12333008, 12422305, and 12373037), the National Key R\&D Program of China (No.
2021YFA1600403/1), the International Centre of Supernovae, Yunnan Key Laboratory (No. 202302AN360001), and the New Cornerstone Science Foundationthrough the XPLORER PRIZE.
JL is supported by the Yunnan Fundamental Research Projects (YFRP) grant No. 202501CF070016 and the Young Talent Project of Yunnan Revitalization Talent Support Program. HC is supported by the CAS "Light of West China", the Young Talent Project of Yunnan Revitalization Talent Support Program, and the Yunnan Fundamental Research Projects (No. 202401BC070007, 202601CJ070008).

Guoshoujing Telescope (the Large Sky Area Multi-Object Fiber Spectroscopic Telescope, LAMOST) is a National Major Scientific Project built by CAS. Funding for the project has been provided by the National Development and Reform Commission. LAMOST is operated and managed by the National Astronomical Observatories, CASs. 

This paper includes data collected by the TESS mission. Funding for the TESS mission is provided by the NASA Explorer Program. This work presents results from the European Space Agency (ESA) space mission Gaia. Gaia data are being processed by the Gaia Data Processing and Analysis Consortium (DPAC). Funding for the DPAC is provided by national institutions, in particular the institutions participating in the Gaia Multi-Lateral Agreement (MLA). The Gaia mission website is \url{https://www.cosmos.esa.int/gaia}. The Gaia archive website is \url{https://archives.esac.esa.int/gaia}.

This work used data of observations with eROSITA telescope onboard SRG observatory. The SRG observatory was built by Roskosmos in the interests of the Russian Academy of Sciences represented by its Space Research Institute (IKI) in the framework of the Russian Federal Space Program, with the participation of the Deutsches Zentrum für Luft- und Raumfahrt (DLR). The SRG/eROSITA X-ray telescope was built by a consortium of German Institutes led by MPE, and supported by DLR.  The SRG spacecraft was designed, built, launched and is operated by the Lavochkin Association and its subcontractors. The science data are downlinked via the Deep Space Network Antennae in Bear Lakes, Ussurijsk, and Baykonur, funded by Roskosmos. The eROSITA data used in this work were processed using the eSASS software system developed by the German eROSITA consortium and proprietary data reduction and analysis software developed by the Russian eROSITA Consortium.

We thank the teams and staff of the SIMBAD database \citep{simbad2000A&AS..143....9W} and VizieR catalogue service, operated at CDS, Strasbourg, France, for providing essential astronomical data.

{\it Software or package}: We use standard data analysis tools in Python environments. Specifically, the TESS light curve is extracted and downloaded with \href{https://docs.lightkurve.org/#}{{\tt lightkurve}} \citep{lightkurve2018ascl.soft12013L}.
Light curve modeling is performed with package \href{https://github.com/lidihei/pylcurve}{\tt LCURVE} \citep{Copperwheat2010MNRAS.402.1824C}. 
MCMC is performed with package \href{https://emcee.readthedocs.io/en/stable/index.html}{\tt emcee} \citep{emcee2013ascl.soft03002F} 3.1.3. 
The binary orbital radial velocity is calculated by \href{https://radvel.readthedocs.io/en/latest/}{\tt radvel} \citep{RadVel_Fulton2018}.
The Python packages \href{https://www.astropy.org/}{\tt astropy} \citep{astropy:2013, astropy:2018, 2022ApJ...935..167A}, \href{https://numpy.org/}{\tt numpy} \citep{Numpy2011CSE....13b..22V, Numpy2020Natur.585..357H} and \href{https://scipy.org/}{\tt scipy} \citep{scipy_jones:_2001, scipy-NMeth2020} are also used.
The HRIES and is extracted by \href{https://iraf-community.github.io/}{{\tt IRAF}} \citep{iraf1986SPIE..627..733T, iraf1993ASPC...52..173T}, 
the DBSP spectra is extracted by \href{https://github.com/lidihei/pyexspec}{{\tt pyexspec}}, their radial velocity is measured using package \href{https://github.com/hypergravity/laspec}{{\tt Laspec}} \citep{Zhangbo2020ApJS..246....9Z}. The SED fitting is performed by using \href{https://speedyfit.readthedocs.io/en/stable/}{\tt SPEEDYFIT}. 
The binary evolutionary model is carried out with \href{https://docs.mesastar.org/en/latest/news/2020-03-05-r12778.html}{\textsc{MESA}} 12778 \citep{Paxton11}.

\section*{Data Availability}

The TESS light curves underlying this article are publicly available at the Mikulski Archive for Space Telescopes (MAST; \url{https://archive.stsci.edu/}), and can be accessed via the MIT Quick Look Pipeline. The ZTF photometric data are available through the Zwicky Transient Facility archive (\url{https://irsa.ipac.caltech.edu/Missions/ztf.html}). Astrometric and photometric data from Gaia DR3 and Gaia EDR3 are publicly available at the Gaia archive (\url{https://archives.esac.esa.int/gaia}). The GALEX, APASS, JOHNSON, 2MASS, and ALLWISE photometry used in the SED fitting are available through their respective public archives.

The reduced 1D spectra and radial velocity measurements (Table~1) will be made available on reasonable request to the corresponding author. The raw spectroscopic data are available at the Keck Observatory Archive (\url{https://koa.ipac.caltech.edu/}) and will be shared on reasonable request for the P200/DBSP and X216/BFOSC observations. The FAST radio data are available on reasonable request to the corresponding author.

The stellar evolution models were computed using \textsc{mesa} version 12778 \citep{Paxton11}, and the input files (inlists) are available on reasonable request to the corresponding author.
 



\bibliographystyle{mnras}
\bibliography{main} 




\appendix

\section{Members of NGC 6397 and 6752}\label{sec:GC_members}
In order to view the positions of BHBs in {\it Gaia} $G$, $G_{\rm BP}-G_{\rm RP}$ CMD, we choose to use the globular clusters NGC 6397 and 6752. We first obtained the center coordinates ($\alpha_{\rm GC}$, $\delta_{\rm GC}$), radius of Gaia-detected cluster members ($R_{0_{\rm GC}}$), parallax ($\varpi_{\rm GC}$), and proper motions ($\mu_{\alpha_{\rm GC}}$, $\mu_{\delta_{\rm GC}}$) from Table A1 of \cite{Vasiliev2021MNRAS.505.5978V}. The members of the two globular clusters are simply selected from {\it Gaia} DR3 using the following criteria,
\begin{itemize}
    \item the angular distances to the globular cluster centers $\le R_{0_{\rm GC}}$
    \item $RUWE \le 1.15$ 
    \item ipd\_frac\_multi\_peak $\le$ 2
    \item ipd\_gof\_harmonic\_amplitude $\le \exp{(0.18(G - 33))}$
    \item visibility\_periods\_used $\ge$ 10 
    \item $\sqrt{(\mu_{\alpha}- \mu_{\alpha_{\rm GC}})^2 + (\mu_{\delta}-\mu_{\delta_{\rm GC}})^2} \le 25$ mas\,yr$^{-1}$
    \item $ |\varpi -\varpi_{\rm GC}| \le 0.15$ mas
    \item parallax\_over\_error $\ge$ 5
    \item $|{\rm phot\_bp\_rp\_excess\_factor} - f(G_{\rm BP}-G_{\rm RP})| \le 3\sigma_{C^*}(G)$
\end{itemize}
where ${\rm phot\_bp\_rp\_excess\_factor} - f(G_{\rm BP}-G_{\rm RP})$ and $\sigma_{C^*}(G)$ are the corrected BP and RP flux excess factor and the corresponding 1$\sigma$ scatter, respectively (details see Eq. (6) and (18) of Riello et al. (2021)\cite{Riello2021A&A...649A...3R}).

\begin{figure*}
    \centering
    \includegraphics[width=\textwidth]{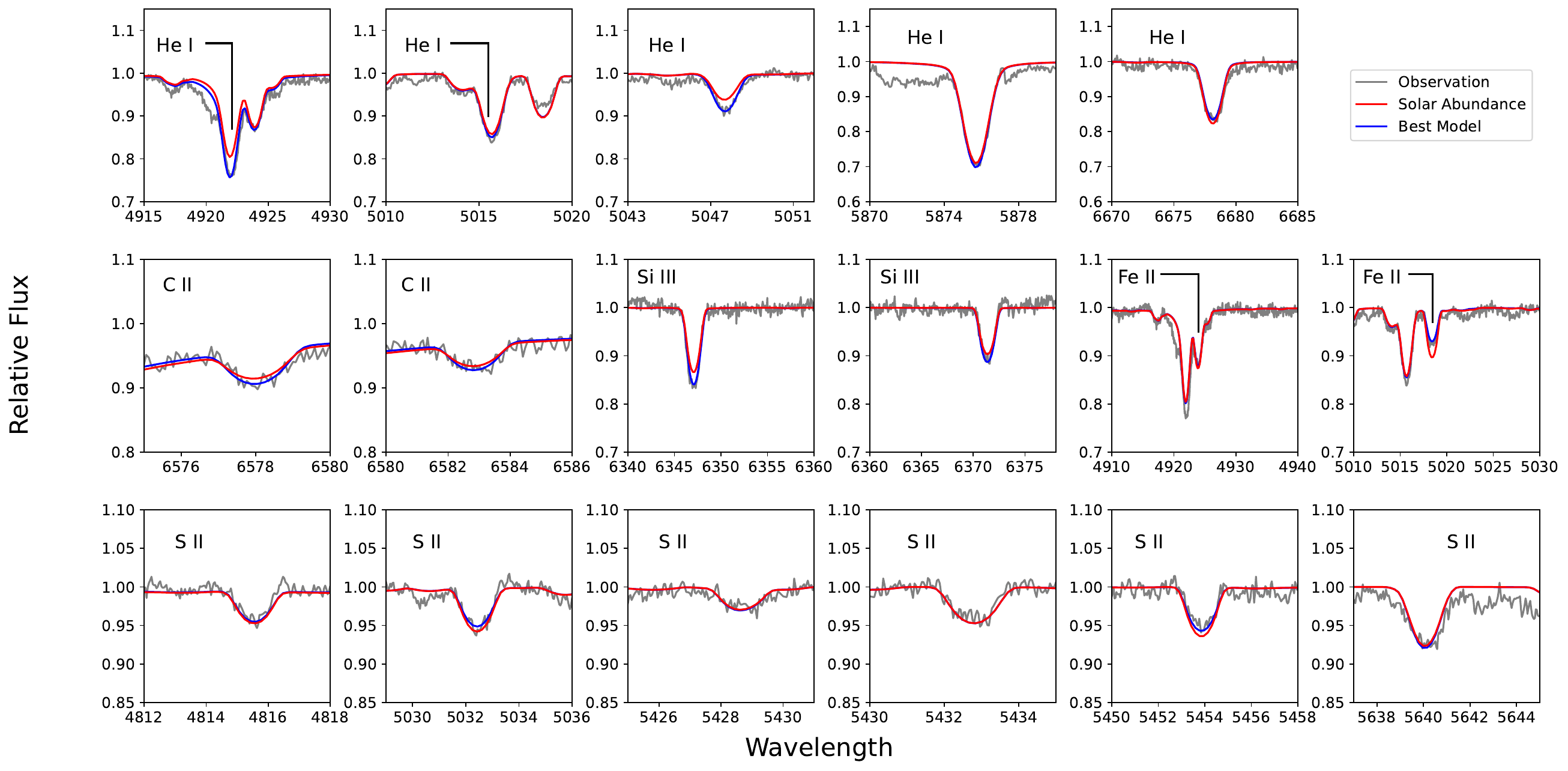}
    \caption{Examples of the best fits for the theoretical results (blue solid lines) to the high-resolution spectra (gray solid lines) for Feige 64. Observed line profiles for lines He\,I, C\,II, Si\,III, Fe\,II and S\,II are superposed on non-LTE synthetic spectra (blue solid lines). The red solid lines represent the synthetic spectra at solar abundance.}
    \label{fig:Compare}
\end{figure*}

\begin{figure}
    \centering
    \includegraphics[width=0.8\columnwidth]{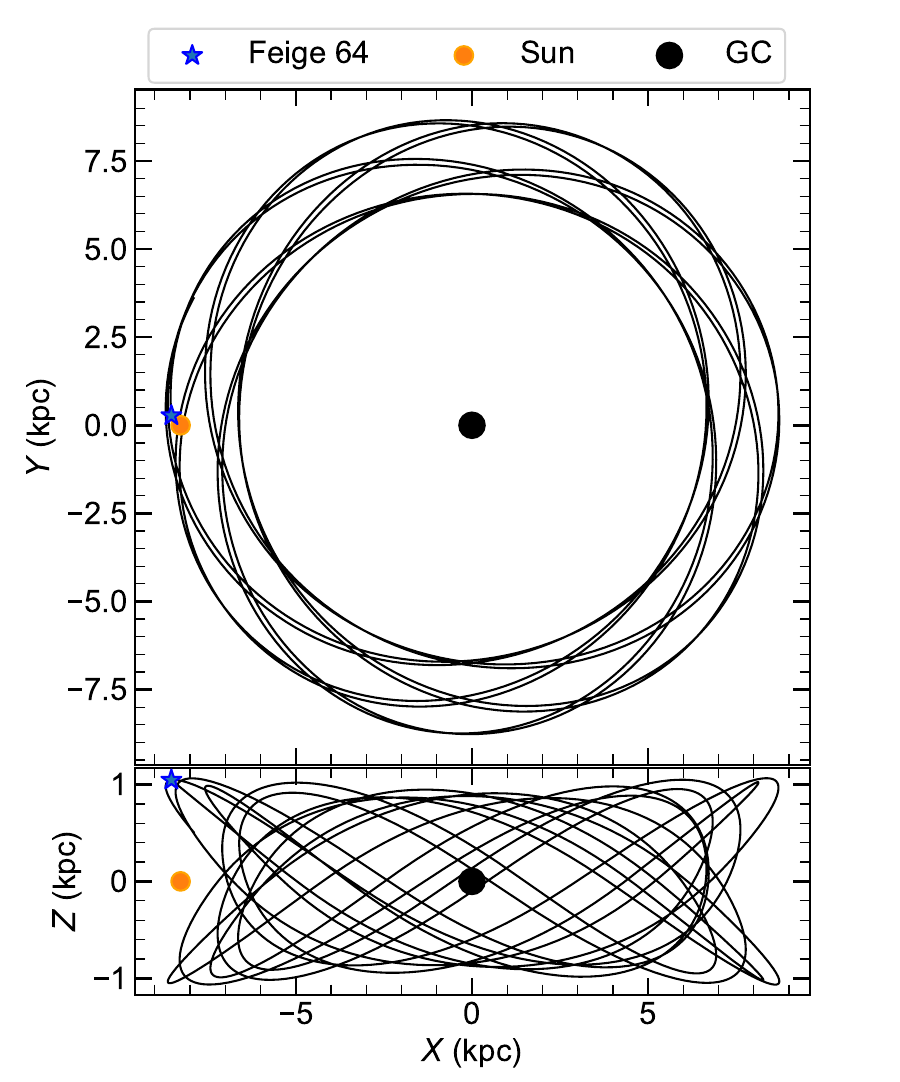}
    \caption{The integrated trajectory of \thestar, in the Galaxy over 2 Gyr, calculated with a time step of 0.1 Myr. The blue star, orange dot and black dot represent the current positions of \thestar, the Sun and the Galactic center, respectively.  }
    \label{fig:space_trajectory}
\end{figure}

\begin{table}
    \centering
    \begin{tabular}{c|c c c }
    \hline
    UT shut& $T_{\rm eff}$ & $\log g$ & $Z$\\
    yyyy-mm-dd hh:mm:ss & K & (cgs) & $Z_{\odot}$\\
    \hline
    2020-01-13T10:35:43 & $15561_{-28}^{+33}$ & $4.1105_{-0.0051}^{+0.0051}$ & $1.669_{-0.061}^{+0.055}$ \\
    2020-01-13T10:35:43 & $15536_{-31}^{+27}$ & $4.1037_{-0.0088}^{+0.0042}$ & $1.593_{-0.047}^{+0.036}$ \\
    2020-01-13T11:16:27&$15475_{-22}^{+20}$ & $4.0982_{-0.0063}^{+0.0056}$ & $1.747_{-0.038}^{+0.038}$\\
    \hline
    mean value & $15524\pm85$ & $4.104\pm0.012$ & $1.67\pm0.15$\\
    \hline
    \end{tabular}
    \caption{The atmosphere parameters derived from DBSP spectra of \thestar. For each parameter, the error of the mean value is the difference between the maximum and minimum values.}
    \label{tab:spec_atom}
\end{table}

\begin{table}
    \centering
    \begin{tabular}{cccc}
    \hline
    $\lambda$\, (\AA) &log X/H+12& $\sigma_{\rm A}$ $^{\rm a}$ &$\sigma_{\rm B}$$^{\rm b}$ \\
    \cline{2-4}
    & \multicolumn{3}{c}{(dex)}\\
    \hline
    \ion{He}{i}\,4921.93 &11.32 & \multirow{5}{*}{0.23}& \multirow{5}{*}{0.1}\\
    \ion{He}{i}\,5015.68 &11.04 & & \\
    \ion{He}{i}\,5047.74 &11.38 & & \\
    \ion{He}{i}\,5875.60 &10.88 & & \\
    \ion{He}{i}\,6678.15 &10.94 & & \\
    \hline
    \ion{C}{ii}\,6578.05 & 8.68 & \multirow{2}{*}{0.02} & \multirow{2}{*}{0.09}\\
    \ion{C}{ii}\,6583.88 &8.65 & & \\
    \hline
    \ion{Si}{ii}\,6347.11 & 7.86 &\multirow{2}{*}{0.08}&\multirow{2}{*}{0.21} \\
    \ion{Si}{ii}\,6371.37 & 7.75 & & \\
    \hline
    \ion{S}{ii}\,4815.55  & 7.28 &\multirow{6}{*}{0.06} &\multirow{6}{*}{0.09} \\
    \ion{S}{ii}\,5032.45  & 7.20 & & \\
    \ion{S}{ii}\,5428.66  & 7.32 & & \\
    \ion{S}{ii}\,5432.80  & 7.30 & & \\
    \ion{S}{ii}\,5453.86  & 7.18 & & \\
    \ion{S}{ii}\,5639.97  & 7.32 & & \\
    \hline
    \ion{Fe}{ii}\,4923.92 & 7.46 & \multirow{2}{*}{0.18}& \multirow{2}{*}{0.08}\\
    \ion{Fe}{ii}\,5018.44 & 7.20 & & \\
    \hline
    \multicolumn{4}{l}{$^{\rm a}$: $\sigma_{\rm A}$ represents the standard deviation of elemental abundances.}\\
    \multicolumn{4}{l}{$^{\rm b}$: $\sigma_{\rm B}$ represents the abundance variation caused by a $\rm$3 \kms{} }\\
    \multicolumn{4}{l}{change in the microturbulent velocity.}
    \end{tabular}
    \caption{Spectral line analysis of Feige 64.}
    \label{tab:specabundace}
\end{table}


\bsp	
\label{lastpage}
\end{document}